\documentclass[onecolumn,groupedaddress,aps,pra,preprint,amsmath,amssymb,showpacs]{revtex4-1}
\usepackage{graphicx,amsmath,amssymb,epsfig,color}
\usepackage{bm}
\usepackage[colorlinks,	linkcolor=red,	anchorcolor=blue,citecolor=blue,bookmarksnumbered]{hyperref}
\allowdisplaybreaks
\begin{document}
\title{Fast-Responding Property of Electromagnetically Induced Transparency in Rydberg Atoms}
\author{Qi Zhang$^{1}$, Zhengyang Bai$^{1}$, and Guoxiang Huang$^{1,2,}$\footnote{gxhuang@phy.ecnu.edu.cn}}
\affiliation{$^1$State Key Laboratory of Precision Spectroscopy,
                 East China Normal University, Shanghai 200062, China\\
             $^2$NYU-ECNU Joint Institute of Physics at NYU-Shanghai, Shanghai 200062, China
             }
\date{\today}

\begin{abstract}
We investigate the transient optical response property of an electromagnetically induced transparency (EIT) in a cold Rydberg atomic gas. We show that both the transient behavior and the steady-state EIT spectrum of the system depend strongly on Rydberg interaction. Especially, the response speed of the Rydberg-EIT can be five-times faster (and even higher) than the conventional EIT without the Rydberg interaction. For comparison, two different theoretical approaches (i.e. two-atom model and many-atom model) are considered, revealing that Rydberg blockade effect plays a significant role for increasing the response speed of the Rydberg-EIT. The fast-responding Rydberg-EIT by using the strong, tunable Rydberg interaction uncovered here is not only helpful for enhancing the understanding of the many-body dynamics of Rydberg atoms but also useful for practical applications in quantum information processing %(e.g. photonic switching and transistors, quantum phase gates, etc.)
by using Rydberg atoms.
\end{abstract}


\maketitle

%%%%%%%%%%%%%%%%%%%%%%%%%%%%%%%%%%%%%%%%%%%%%%%%%%%%%%%%%%%%%%%
\section{Introduction}
In the past two decades, much attention has been paid to the research of cold Rydberg atomic gases~\cite{And,Mou,Saffman,Low,REITreview,Fir,Mur}, i.e. highly excited atoms with large principal quantum number~\cite{Gallagher} working under an ultracold environment. Due to their exaggerated properties, including long lifetime, large electric dipole moment, strong and controllable atom-atom interaction (called Rydberg interaction for short), etc., Rydberg atoms have promising applications in quantum calculating and quantum information, precision spectroscopy and precision measurement, manipulation and simulation of quantum many-body states, and so on~\cite{Saffman,Low,REITreview,Fir,Mur}.

Since the pioneering theoretical and experimental works carried out by Friedler {\it et al.}~\cite{Fri} and by
Mohapatra {\it et al.}~\cite{Mohapatra}, in recent years considerable interest has been focused on the electromagnetically induced transparency (EIT) in Rydberg atomic gases (see Refs.~\cite{REITreview,Fir,Mur} for details). EIT is a typical quantum interference effect in three-level atoms induced by a control laser field, by which the absorption of a probe field can be significantly suppressed. Light propagation in EIT systems displays many striking features, which include (in addition to the significant suppression of light absorption) large reduction of group velocity, giant enhancement of Kerr nonlinearity, etc.~\cite{EITreview} Rydberg-EIT has important applications, such as direct and non-destructive coherent optical detection~\cite{Mohapatra}, design and fabrication of devices in quantum information processing (e.g., all-optical switches and transistors) at single-photon level~\cite{Gor0,Baur,Gorn1,Tia1,Gorn2,Tia2,Kuzmich}, and development of quantum nonlinear optics in correlated quantum many-body systems with strong driving and dissipation outside of equilibrium~\cite{REITreview,Fir,Mur,Carr,Marc}.

However, up to now most studies on Rydberg-EIT are limited to the steady-state property or long-time behavior, in which the transient response process [appearing when the control (or probe) field is switched on] was not taken into account. For many practical applications, such as the performance of all-optical switches and transistors, the response speed of Rydberg-EIT is vital. Thus it is very necessary to explore the transient optical response of Rydberg-EIT, which is important not only for the understanding of the physical property of EIT in Rydberg atoms, but also for practical applications of all-optical switches and transistors, and even general quantum memory processes based on Rydberg-EIT~\cite{Gor0,Baur,Gorn1,Tia1,Gorn2,Tia2,Kuzmich}.

In this work, we investigate, both analytically and numerically, the transient optical response property of an Rydberg-EIT when the control field in the system is switched on from zero into a finite value. We shall show that both the transient-state behavior and the steady-state EIT spectrum of the Rydberg atomic gas depend strongly on Rydberg interaction. In particular, the response speed of the Rydberg-EIT can be five-times faster than the conventional EIT without the Rydberg interaction, and may be increased further if the system parameters are optimized. For comparison, two different theoretical models are considered, i.e., a two-atom model
for which the equation of motion of the density matrix of the system is solved exactly by using a numerical calculation, and a many-atom model for which equations of motion of reduced density matrix
(i.e. many-body correlators) are solved by using an approach beyond mean-field approximation.
Two models give consistent results, which show that Rydberg blockade effect plays a significant role for increasing the response speed of the Rydberg-EIT. The fast-responding Rydberg-EIT by using the strong, tunable Rydberg interaction found here is not only helpful for enhancing the understanding of the many-body dynamics of Rydberg atoms but also useful for practical applications in quantum information processing by using Rydberg atoms.

The paper is arranged as follows. In Sec.~\ref{sec2}, we describe the two-atom model and give numerical results of the transient response and the estimation on the response time of the Rydberg-EIT. In Sec.~\ref{sec3}, we introduce the many-atom model, present analytical results of the transient response by using an approach of reduced density approach, and make a comparison with the result obtained from the two-atom model. Finally, in Sec.~\ref{sec4} we give a discussion and a summary of the results obtained in this work. The information about calculation details of the main text are presented in Appendixes.

\section{Transient response of the Rydberg-EIT: two-atom model}\label{sec2}
\subsection{Two-atom model}~\label{sec2a}
\begin{figure}
\includegraphics[ width=0.8\columnwidth]{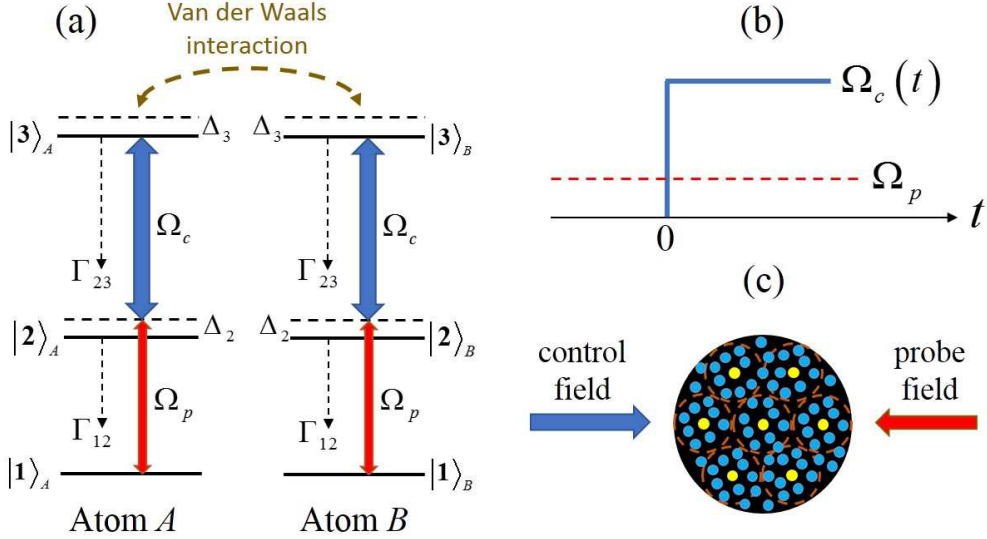}
\caption{\footnotesize (Color online) (a)~Level configuration and excitation scheme of the two-atom model, which consists of two identical atoms, $A$ and $B$, with three internal states $|1\rangle_l$, $|2\rangle_l$, and $|3\rangle_l$ (Rydberg state), interacting via van der Waals (Rydberg) interaction. $\Gamma_{12}$ ($\Gamma_{23}$): decay rate from $|2\rangle_l$ to $|1\rangle_l$ (from $|3\rangle_l$ to $|2\rangle_l$); $\Delta_{2}$
($\Delta_{3}$): one- (two-) photon detuning; $\Omega_{p}$ ($\Omega_{c}$): half Rabi frequency of the probe (control) field coupling to the transition $|1\rangle_l\leftrightarrow|2\rangle_l$ ($|2\rangle_l\leftrightarrow|3\rangle_l$) ($l=A,B$). (b)~Time sequence for the probe (red dashed line) and the control (blue solid line) fields.
(c)~Schematic of Rydberg blockade in the many-atom model. The Rydberg interaction between atoms blocks the excitation of the atoms within blockade spheres (i.e. the ones with the boundary marked by the orange dashed lines). In each blocked sphere only one Rydberg atom (small yellow sphere) is excited and excitations of other atoms (small blue spheres) to their Rydberg states are suppressed.
}
\label{Fig1}
\end{figure}

Firstly, we consider a system consisting of only two identical atoms, $A$ and $B$, with three internal states
driven by two laser fields [Fig.~\ref{Fig1}(a)].
One of them is a probe field, which has the center angular frequency $\omega_{p}$ (half Rabi frequency $\Omega_{p}$) and couples to the transition between ground state $|1\rangle_l$ and excited (intermediate) state $|2\rangle_l$; another is a control field, which has center angular frequency $\omega_{c}$ (half Rabi frequency $\Omega_{c}$) and couples to the transition between the state $|2\rangle_l$ to Rydberg state $|3\rangle_l$. $\Gamma_{12}$ ($\Gamma_{23}$) is the decay rate from the excited state to the ground state (from the Rydberg state to the excited state), $\Delta_{2}=\omega_{p}-(\omega_{2}-\omega_{1})$  [$\Delta_{3}=(\omega_{p}+\omega_{c})-(\omega_{3}-\omega_{1})$]
is one-photon (two-photon) detuning, with $\hbar\omega_{\alpha}$ the eigenenergy of the state $|\alpha\rangle$.
For simplicity, the Rydberg states $|3\rangle_l$  ($l=A$, $B$) is assumed to be $|nS_{1/2}\rangle$ (with $n$ principle quantum number). There is a long-range van der Waals (Rydberg) interaction between the Rydberg states $|3\rangle_A$ and $|3\rangle_B$.

Under electric-dipole approximation, the Hamiltonian of the system is given by ${\hat H}={\hat H}_{A}+{\hat H}_{B}+{\hat H}_{AB}$. Here ${\hat H}_{A}$ (${\hat H}_{B}$) is the single-atom Hamiltonian for atom $A$ (atom $B$), and ${\hat H}_{AB}$ is the van der Waals (vdW) interaction between two atoms. Under rotating-wave approximation, the Hamiltonian in interaction picture reads
\begin{equation}
{\hat H}=-\hbar\sum_{l=A,B}\left[\sum_{\alpha=1}^{3}\Delta_{\alpha}{\hat\sigma}_{\alpha\alpha}^{l}+\left(\Omega_{p}{\hat\sigma}_{21}^{l}+\Omega_{c}{\hat\sigma}_{32}^{l}+{\rm H.c.}\right)\right]+\hbar{\hat\sigma}_{33}^{A}V_{AB}{\hat\sigma}_{33}^{B},
\label{Schrodinger Hamiltonian}
\end{equation}
where ${\hat\sigma}_{\alpha\beta}^{l}\equiv|\alpha\rangle_{l\, l}\langle\beta|$ is the transition operator of atom $l$ ($l=A,\,B$),
$\Omega_{p(c)}=[{\bf e}_{p(c)}\cdot{\bf p}_{21(32)}]{\cal E}_{p(c)}/\hbar$ is the half Rabi frequency of the probe (control) field (with ${\bf p}_{\alpha\beta}$ the electric dipole matrix element associated with the transition from $|\beta\rangle$ to $|\alpha\rangle$), and $V_{AB}=-C_{6}/r_{AB}^6$ is the vdW interaction potential (with $r_{AB}\equiv |{\bf r}_{A}-{\bf r}_{B}|$ the separation between atom $A$ and atom $B$ and $C_{6}$ the dispersion coefficient approximately scaling as $n^{11}$).

The state vector of the system in the interaction picture is $|\Psi\rangle=\sum_{\alpha,\mu=1}^3 a_{\alpha\mu}|\alpha\mu\rangle$, with $|\alpha\mu\rangle\equiv|\alpha\rangle_A\,|\mu\rangle_B$ and $a_{\alpha\nu}$ the corresponding probability amplitude. The density matrix of the system, defined by $\hat{\rho}\equiv |\Psi\rangle\langle \Psi|$, reads
\begin{equation}\label{EQ8}
\hat{\rho}=\sum_{\alpha,\beta=1}^3 \sum_{\mu,\nu=1}^3\rho_{\alpha\beta,\mu\nu}|\alpha\mu\rangle\langle\beta\nu|=\sum_{\alpha,\beta=1}^3\sum_{\mu,\nu=1}^3\rho_{\alpha\beta,\mu\nu}\hat{\sigma}_{\alpha\beta}^A\,\hat{\sigma}_{\mu\nu}^B,
\end{equation}
where $\rho_{\alpha\beta,\mu\nu}\equiv \langle\alpha\mu|\hat\rho|\beta\nu\rangle=a_{\alpha\mu}a^{\ast}_{\beta\nu}$
satisfying $\sum_{\alpha\mu=1}^{3}\rho_{\alpha\alpha,\mu\mu}=1$ and  $\rho_{\alpha\beta,\mu\nu}^{\ast}=\rho_{\beta\alpha,\nu\mu}$.
The master equation governing the evolution of the density matrix reads
\begin{equation}
i\hbar\frac{\partial{\hat\rho}}{\partial t}=\left[{\hat H},{\hat\rho}\right]+\Gamma{\hat\rho},
\label{Schrodinger Eqs}
\end{equation}
where $\Gamma$ is a $9\times9$ relaxation matrix representing the decay rates due to spontaneous emission and dephasing in the system. An explicit form of the master equation is presented in Eq.~(\ref{Two-atom Eqs}) of {\color{blue}Appendix~\ref{AppAa}}.

The reduced one-atom density matrix $\rho^{A}$ for atom $A$ is given by ${\hat\rho}^{A}={\rm Tr}^{B}(\hat\rho)$~\cite{Gillet}, i.e. the partial trace of the density matrix over atom B. Then it is easy to show that
\begin{equation}\label{RhoA}
\rho_{\alpha\beta}^{A}=\sum_{\mu=1}^{3}\rho_{\alpha\beta,\mu\mu}.
\end{equation}
Similarly, the reduced one-atom density matrix $\rho^{B}$ for atom $B$ is given by ${\hat\rho}^{B}={\rm Tr}^{A}(\hat\rho)$, and we have  $\rho_{\mu\nu}^{B}=\sum_{\alpha=1}^{3}\rho_{\alpha\alpha,\mu\nu}$.
Note that, due to the symmetry of the Hamiltonian Eq.~(\ref{Schrodinger Hamiltonian}) by exchanging atom $A$ and atom $B$, one has numerically $\rho_{\alpha\beta}^{A}=\rho_{\alpha\beta}^{B}$; in addition, for very large atomic separation ($r_{AB}\rightarrow\infty$), $V_{AB}\rightarrow 0$, and hence we have $\rho_{\alpha\beta,\mu\nu}=\rho_{\alpha\beta}^{A}\rho_{\mu\nu}^{B}$. In this situation, the system is reduced into two independent atoms and hence the Rydberg-EIT becomes a conventional one without atomic interaction.

The physical system described in the present work can be easily realized by experiment. One of candidates is $^{87}$Rb atoms trapped in a microtrap, with the atomic states [shown in Fig.~\ref{Fig1}(a)] assigned as
$|1\rangle=|5s^2S_{1/2},F=2\rangle$, $|2\rangle=|5p^2P_{3/2},F=3\rangle$, and $|3\rangle=|60s^2S_{1/2}\rangle$, with $\Gamma_{12}=2\pi\times6\,{\rm MHz}$, $\Gamma_{23}=1\,{\rm kHz}$, $C_{6}=-2\pi\times140\,{\rm GHz\cdot\mu m^6}$ for $n=60$~\cite{Pritchard,Steck}. In this work, as done by Li and Xiao~\cite{Xiao}, we consider the transient optical response of the Rydberg-EIT by using the time sequence shown in Fig.~\ref{Fig1}(b). That is to say, when $t<0$ the probe field is present but with no control field applied, so the system has an optical response of a typical two-level atomic system; at $t=0$ the control field is rapidly switched on; for $t>0$ the system displays a transient optical response process, till to the establishment of a steady-state Rydberg-EIT at some time $T_{R}$ (i.e. the response time of the Rydberg-EIT; see below).

\subsection{Transient response property of the Rydberg-EIT in the two-atom model}~\label{sec2b}

Since for $t<0$ the control field is absent (i.e. $\Omega_{c}=0$), the Rydberg states are empty and the Rydberg interaction plays no role. Thus the system performs as two independent atoms with the ground and excited states coupled by the probe field. Then, if taking $t=0$ as an initial time, the initial condition of the system is given by $\rho_{\alpha\beta,\mu\nu}(t)|_{t=0}\equiv
\rho_{\alpha\beta,\mu\nu}(0)=\rho_{\alpha\beta}^{A}(0)\rho_{\mu\nu}^{B}(0)\,(\alpha,\beta,\mu,\nu=1,2)$, with other $\rho_{\alpha\beta,\mu\nu}(0)=0$. Here
$\rho_{11}^{A}(0)=\rho_{11}^{B}(0)=
[2\gamma_{21}|\Omega_{p}|^2+\Gamma_{12}|d_{21}|^2]/D$,
$\rho_{22}^{A}(0)=\rho_{22}^{B}(0)=
\Gamma_{12}|d_{21}|^2/D$,
$\rho_{21}^{A}(0)=\rho_{21}^{B}(0)=
-d_{21}^{\ast}\Gamma_{12}\Omega_{p}/D$,
with $D=4\gamma_{21}|\Omega_{p}|^2+\Gamma_{12}|d_{21}|^2$, and $d_{21}=\Delta_{2}+i\gamma_{21}$  ($\gamma_{21}=\Gamma_{12}/2$  is the dephasing rate between $|1\rangle_{l}$ and $|2\rangle_{l}$; $l=A, B$).
The dynamical behavior of the system when the control field is switched on can be obtained through solving
the equation of motion of the two-atom density matrix Eq.~(\ref{Schrodinger Eqs}) by using the well-known standard Runge-Kutta method under the initial condition given above.

We are interested in the transient optical response of the Rydberg-EIT, which can be described by the time evolution of the optical susceptibility $\chi_p (t)$ of the probe field, proportional to the one-atom coherence $\rho_{21}^{A}$ (or $\rho_{21}^{B}$). From Eq.~(\ref{RhoA}) we have
\begin{equation}
\rho_{21}^{A}(t)=\rho_{21,11}(t)+\rho_{21,22}(t)+\rho_{21,33}(t),
\label{sgm21}
\end{equation}
and similarly $\rho_{21}^{B}(t)=\rho_{11,21}(t)+\rho_{22,21}(t)+\rho_{33,21}(t)$, which is equal to $\rho_{21}^{A}(t)$.

\begin{figure}
\includegraphics[width=1.0\columnwidth]{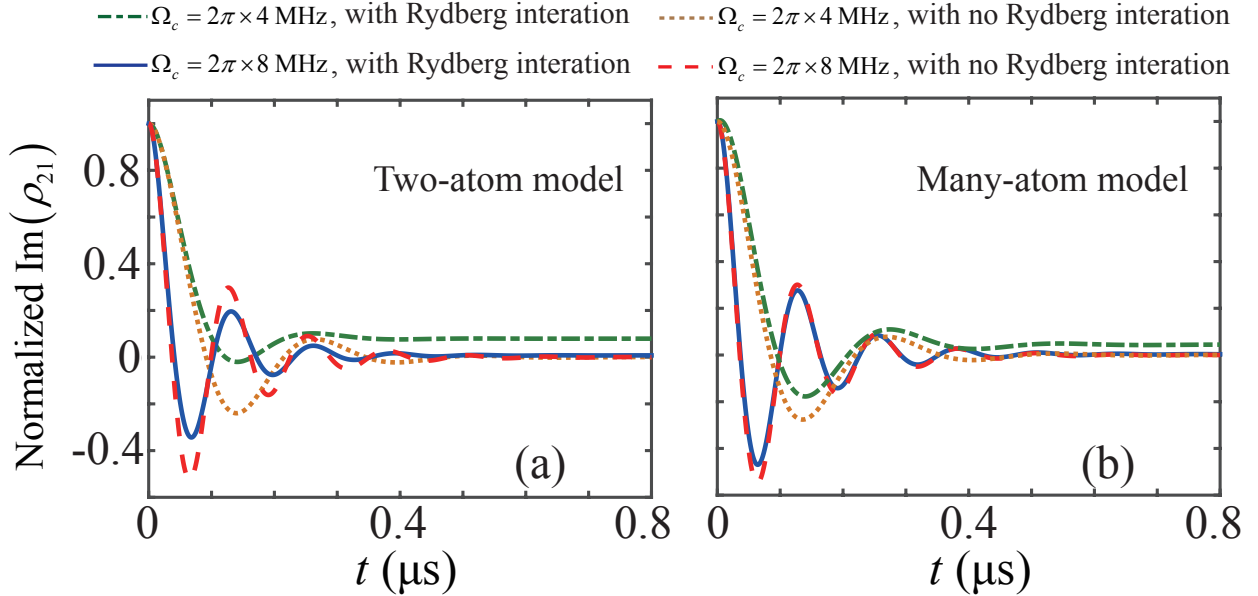}
\caption{\footnotesize  (Color online) Transient response behavior of the Rydberg-EIT as a function of time $t$.
(a)~Normalized absorption of the two-atom model for $\Omega_{p}=0.2\Gamma_{12}$, characterized by ${\rm Im}(\rho_{21})$  [$\rho_{21}(t)\equiv \rho_{21}^A(t)]$. The green dashed-dotted line is for   $\Omega_{c}=2\pi\times4\,{\rm MHz}$ with the Rydberg interaction [$V_{AB}=1\,{\rm GHz}$ ($r_{AB}=3.10\,{\rm\mu m}$)]; the brown dotted line is for  $\Omega_{c}=2\pi\times4\,{\rm MHz}$  with no Rydberg interaction ($V_{AB}=0$).
(b)~${\rm Im}(\rho_{21})$ of the many-atom model for $\Omega_{p}=0.05\Gamma_{12}$ as a function of $t$.  The green dashed-dotted line is for $\Omega_{c}=2\pi\times4\,{\rm MHz}$ with a high atomic density ($N_{a}=1.2\times10^{10}\,{\rm cm}^{-3}$) and hence significant Rydberg interaction; the brown dotted line is for $\Omega_{c}=2\pi\times4\,{\rm MHz}$ with a low atomic density ($N_{a}=1\times10^{8}\,{\rm cm}^{-3}$) and hence negligible Rydberg interaction.
Results for a large control field, i.e.  $\Omega_{c}=2\pi\times8\,{\rm MHz}$, are also shown. In both panels, blue solid lines and red dashed lines are for the case with significant Rydberg interaction and the case with negligible Rydberg interaction, respectively.
}
\label{Fig2}
\end{figure}

Shown in Fig.~\ref{Fig2}(a)
is the numerical result on the transient response behavior of the Rydberg-EIT of the two-atom model as a function of time $t$ for $\Delta_2=\Delta_3=0$, characterized by the normalized absorption ${\rm Im}(\rho_{21})$, i.e. the imaginary part of  $\rho_{21}(t)$  [$\equiv \rho_{21}^A(t)]$, as a function of $t$ for
$\Omega_{p}=0.2\Gamma_{12}$. The green dashed-dotted line is for the case $\Omega_{c}=2\pi\times4\,{\rm MHz}$ with the Rydberg interaction [$V_{AB}=1\,{\rm GHz}$ ($r_{AB}=3.10\,{\rm\mu m}$)]; the brown dotted line is for the case $\Omega_{c}=2\pi\times4\,{\rm MHz}$ with no Rydberg interaction ($V_{AB}=0$).

From the figure we see that:
(i)~As the control field is switched on at $t=0$, both the absorption curves of the EIT with and without the Rydberg interaction display a damped
%in-phase
oscillation. For large $t$, the absorption is increased and ${\rm Im}(\rho_{21})$ reaches to a small steady-state value. The small absorption at the steady-state is due to the quantum destructive interference effect induced by the strong control field.
(ii)~Comparing with the case with no Rydberg interaction  where a transient gain [i.e. ${\rm Im}(\rho_{21})<0$] may happen (the brown dotted line), the oscillation amplitude for the case with the Rydberg interaction is smaller (the green dashed-dotted line). The reason is that the presence of the Rydberg interaction contributes an out-of-phase impact on the EIT with no Rydberg interaction.  This point can be also seen from Eq.~(\ref{rho21}), obtained by using the many-atom model in the next section.
(iii)~The oscillating frequency of the response curve for large $\Omega_{c}$ is larger than that of small $\Omega_{c}$, regardless of the Rydberg interaction (the red dashed and the blue solid lines).
This is because the coherence property of the system is enhanced when $\Omega_{c}$ becomes larger, resulting in an enhanced oscillation before reaching its steady-state value.

\begin{figure}
\includegraphics[width=0.9\columnwidth]{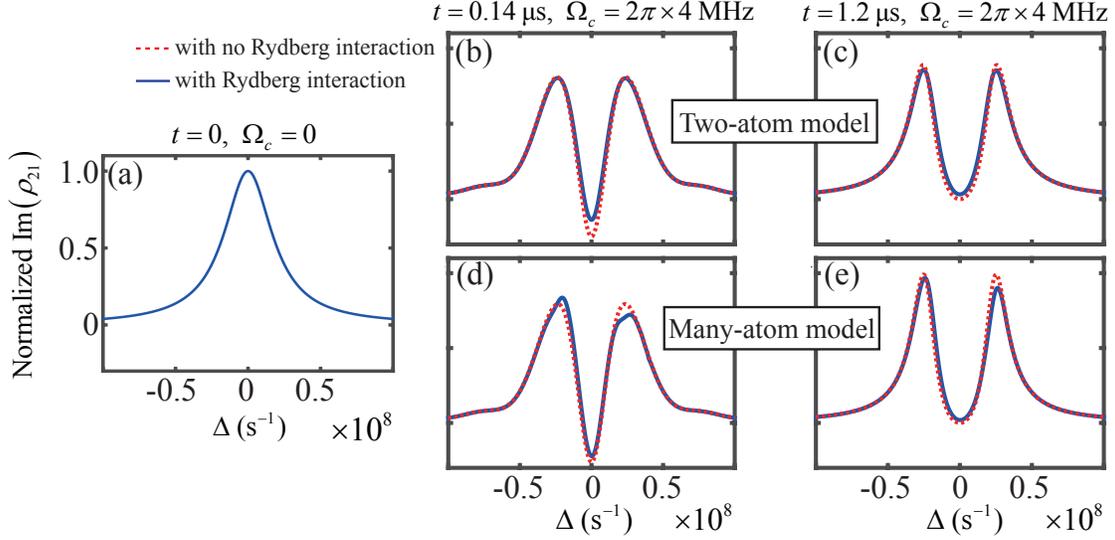}
\caption{\footnotesize  (Color online) Transient response behavior of the Rydberg-EIT
as a function of the probe-field detuning $\Delta\,(\equiv\Delta_{2}=\Delta_{3})$.
(a) Normalized absorption spectrum ${\rm Im}(\rho_{21})$  for $t=0$, where $\Omega_c=0$.
(b)~${\rm Im}(\rho_{21})$ at $t=0.14\,{\rm\mu s}$ for $\Omega_c=2\pi\times 4$\,MHz.
(c)~${\rm Im}(\rho_{21})$ at $t=1.2\,{\rm\mu s}$ for $\Omega_c=2\pi\times 4$\,MHz.
Both (b) and (c) are obtained from the two-atom model with $\Omega_{p}=0.2\Gamma_{12}$, where the blue solid line is the EIT spectrum with the Rydberg interaction $V_{AB}=1\,{\rm GHz}$ ($r_{AB}=3.10\,{\rm\mu m}$), while the red dashed line is the EIT spectrum without the Rydberg interaction ($V_{AB}=0$).
(d) and (e) are respectively the same with (a) and (b), but obtained by the many-atom model
with $\Omega_{p}=0.03\Gamma_{12}$, where the blue solid line is the EIT spectrum with a high atomic density $N_{a}=1.2\times10^{10}\,{\rm cm}^{-3}$ (significant Rydberg interaction), while the red dashed line is the EIT spectrum with a low atomic density $N_{a}=1\times10^{8}\,{\rm cm}^{-3}$ (negligible Rydberg interaction).
}
\label{Fig3}
\end{figure}

In order to seek more information of the character on the Rydberg-EIT, the transient response spectrum of the system as a function of the probe-field detuning $\Delta\,(\equiv\Delta_{2}=\Delta_{3})$ is also calculated, with the result plotted in Fig.~\ref{Fig3}. Fig.~\ref{Fig3}(a)
shows the normalized absorption spectrum ${\rm Im}(\rho_{21})$ for $t=0$, which, due to $\Omega_c=0$,
has only a single peak of Lorentz type, typical for a two-level atom coupled with a laser field.
Shown in Fig.~\ref{Fig3}(b) and Fig.~\ref{Fig3}(c) are respectively results of ${\rm Im}(\rho_{21})$ at $t=0.14\,{\rm\mu s}$ and $t=1.2\,{\rm\mu s}$ for $\Omega_c=2\pi\times 4$\,MHz and $\Omega_{p}=0.2\Gamma_{12}$, where the blue solid line is for the case with the Rydberg interaction $V_{AB}=1\,{\rm GHz}$ ($r_{AB}=3.10\,{\rm\mu m}$) and the red dashed line is for the case without the Rydberg interaction ($V_{AB}=0$).

From the figure we see that:
(i)~When the control field is switched on ($t>0$), the original single-peak absorption spectrum at $t=0$ [Fig.~\ref{Fig3}(a)] evolves into a two-peak structure (i.e. a EIT transparency window is opened near $\Delta=0$) and the separation between the two peaks is gradually increased as $t$ increases [Fig.~\ref{Fig3}(b)-\ref{Fig3}(e)]; in addition, a transient gain [i.e. ${\rm Im}(\rho_{21})<0$] is observed at $t=0.14\,{\rm\mu s}$ before the absorption spectrum reaches to the final steady-state value [Fig.~\ref{Fig3}(b), Fig.~\ref{Fig3}(c)].
(ii)~The depth of the EIT transparency window for the case of the EIT with the Rydberg interaction (blue solid line) is shallower than that of the EIT without Rydberg interaction (red dashed line) [Fig.~\ref{Fig3}(b), Fig.~\ref{Fig3}(c)], which means that, comparing with the EIT with no Rydberg interaction, the absorption in the EIT with the Rydberg interaction is stronger.

The dispersion spectrum of the system is described by the real part of the atomic coherence $\rho_{21}$, i.e. Re($\rho_{21})$, as a function of $\Delta$, which has been shown in the panels (a)-(c) of Fig.~\ref{Fig6} of {\color{blue}Appendix~\ref{AppAb}}. One sees that:
(i)~When the control field is switched on ($t>0$), the dispersion spectrum at $t=0$, which displays
an anomalous dispersion [Fig.~\ref{Fig6}(a)], evolves into one with normal dispersion near $\Delta=0$;
(ii)~Near $\Delta=0$, there is a only small difference of the dispersion behavior between the case with and without the Rydberg interaction.

%%%%%%%%%%%%%%%%%%%%%%%%%%%%%%%%%%%
\begin{table}[!hbp]
\caption{\footnotesize  Response time $T_{R}$ of the Rydberg-EIT for $\Omega_{p}=0.3\Gamma_{12}$ obtained by using the two-atom model. }
\vspace{3mm}
\newcommand{\tabincell}[2]{\begin{tabular}{@{}#1@{}}#2\end{tabular}}
\begin{tabular}{c|c|c}
\hline
$\Omega_{c}$ & \tabincell{c}{\,\,\,$T_R$ {\footnotesize with Rydberg interaction}\,\, \,\\ ($V_{AB}=1\,{\rm GHz}$)} & \tabincell{c}{\,\,\,$T_R$ {\footnotesize with no Rydberg interaction}\,\,\, \\ ($V_{AB}=0$)} \\
\hline
\,\,\,$2\pi\times4\,{\rm MHz}$\,\,\, & $0.26\,{\rm\mu s}$ & $1.40\,{\rm\mu s}$ \\
\hline
\,\,\,$2\pi\times8\,{\rm MHz}$\,\,\, & $0.43\,{\rm\mu s}$ & $1.57\,{\rm\mu s}$ \\
\hline
\end{tabular}
\label{Table}
\end{table}
Based on the above results, we can deduce that the EIT with the Rydberg interaction has a fast response time than the EIT without Rydberg interaction. To support this conclusion, we give a quantitative estimation on the response time of the Rydberg-EIT.
According to engineering control theory~\cite{Ogata,Levine}, the response time $T_R$ of a transient response process
may be defined as the minimum time after which the temporal variation of the response function of the transient response process always keeps within a error range $2\Delta_{\rm err}$ ($\Delta_{\rm err}$ is usually set to $0.05$~\cite{Ogata}). A simple mathematical illustration to explain the concept of the response time of a transient response process is given in {\color{blue}Appendix~\ref{AppAc}}.

Based on the above definition, the response time $T_R$ of the Rydberg-EIT is calculated. Shown in Table~\ref{Table} is the result of $T_R$ for $\Omega_{p}=0.3\Gamma_{12}$, obtained by the two-atom model for different $\Omega_{c}$. From Table~\ref{Table} we have the following conclusions:
(i) The response speed of the Rydberg-EIT can be faster than that of the EIT without Rydberg interaction. Especially, for small control field, the response time of the Rydberg-EIT can be five times smaller than that of the EIT without Rydberg interaction. The physical reason for the fast-responding property of the Rydberg-EIT is due to the Rydberg blockade in the system, where the strong Rydberg interaction shifts the Rydberg state $|3\rangle$ out of resonance, and then blocks its excitation. As a result, atoms nearly remain in their initial two-level atomic states, so that the steady-state of  EIT for the interacting system can be achieved in an early time.
(ii) The response time of EIT grows as $\Omega_c$ increases. The physical reason is that, as $\Omega_{c}$ increases, the oscillation frequency of ${\rm Im}(\rho_{21})$ increases due to the enhancement of the coherence of the system. Thus a longer time is needed for $\rho_{21}$ evolving into steady-state. This point can be clearly seen by the blue solid line and the red dashed line in Fig.~\ref{Fig2}(a). Thus for shortening the response time of EIT, one should make moderate $\Omega_{c}$ (small but still satisfies the EIT condition, i.e., $|\Omega_{c}|^2>\gamma_{21}\gamma_{31}$~\cite{EITreview}).

\begin{figure}
\includegraphics[width=0.55\columnwidth]{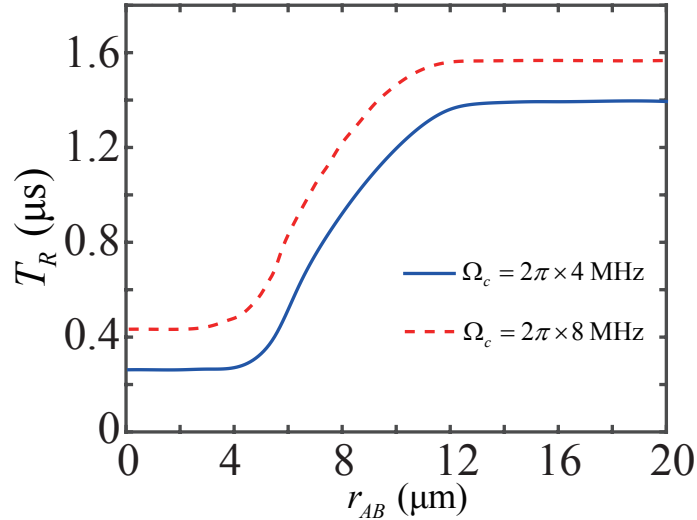}
\caption{\footnotesize  (Color online) The response time $T_R$ of the Rydberg-EIT for as a function of $r_{AB}$, obtained with the two-atom model for $\Omega_{c}=2\pi\times4\,{\rm MHz}$ (blue solid line) and $\Omega_{c}=2\pi\times8\,{\rm MHz}$ (red dashed line), with $\Omega_{p}=0.3\Gamma_{12}$. }
\label{Fig4}
\end{figure}

Note that the response time of the Rydberg-EIT can be changed as the Rydberg interaction is varied. Shown in Fig.~\ref{Fig4}
is the response time $T_{R}$ as a function of $r_{AB}$ in the Rydberg-EIT system obtained by the two-atom model. We see that
$T_{R}$ is shortened as $r_{AB}$ is reduced, which means that one can reduce the Rydberg-EIT response time by increasing the atomic density. However, $T_{R}$ is saturated for small $r_{AB}$. This is due to the effect of ``soft core'', resulted from strong Rydberg blockade effect, where the excitation to Rydberg states is completely blockaded for very closed atoms. Additionally, from the figure
we know that, in general, the response time grows as the control field is increased, regardless the Rydberg interaction.

\section{Transient response of the Rydberg-EIT: many-atom model}\label{sec3}

\subsection{Many-atom model and reduced density matrix approach}\label{sec3a}

In the last section we have shown that the Rydberg-EIT has a fast response speed than conventional EIT without Rydberg interaction. But the result given there is obtained by using a two-atom model, and thus cannot tell us how about the situation if the system contains a large amount of atoms. To answer this question, in this section we investigate the transient response behavior of a many-atom system with Rydberg interaction, for which, however, the density matrix method used in the last section is hard to apply even for a numerical approach since
the size of the Hilbert space is exponentially expanded as the atomic number of the system increases.
Alternatively, here we employ an approach of reduced density matrix~\cite{Schmpp,SevincliPRL,StanojevicTheo,BaiOE} beyond mean-field approximation to solve analytically equations of motion of many-body correlators by a method of multiple-scales~\cite{Newell,Jeffery}.

The Hamiltonian of in a system with $N$ atoms with Rydberg interaction is given by ${\hat H}_{\rm H}(t)=N_{a}\int_{-\infty}^{+\infty}d^3{\bf r}{\hat{\cal H}}_{\rm H}({\bf r},t)$, where $N_{a}$ is atomic density, and ${\hat{\cal H}}_{\rm H}({\bf r},t)$ is the Hamiltonian density, given by~\cite{StanojevicTheo,BaiOE}
\begin{align}
\nonumber {\hat{\cal H}}_{\rm H}({\bf r},t)=
& \hbar\sum_{\alpha=1}^{3}\omega_{\alpha}{\hat S}_{\alpha\alpha}({\bf r},t)-\hbar\left[\Omega_{p}\hat{S}_{12}({\bf r},t)+\Omega_{c}\hat{S}_{23}({\bf r},t)+{\rm H.c.}\right]\\
& +N_{a}\int_{{\bf r}^{\prime}\neq{\bf r}} d^3{\bf r}^{\prime}\hat{S}_{33}({\bf r}^{\prime},t)\hbar V({\bf r}^{\prime}-{\bf r})\hat{S}_{33}({\bf r},t),
\label{Heisenberg Hamiltonian}
\end{align}
where $\hat{S}_{\alpha\beta}({\bf r},t)=|\beta\rangle\langle \alpha|e^{i[({\bf k}_{\beta}-{\bf k}_{\alpha})\cdot{\bf r}-(\omega_{\beta}-\omega_{\alpha}+\Delta_{\beta}-\Delta_{\alpha})t]}\,(\alpha,\beta=1,2,3)$ is the transition operator related to the states $|\alpha\rangle$ and $|\beta\rangle$, the last term on the right hand side is the contribution from the Rydberg interaction, with $\hbar V({\bf r}'-{\bf r})$ the interaction potential between the Rydberg atoms located at the position ${\bf r}$ and ${\bf r}'$.

Due to the Rydberg interaction, the Rydberg excitation of one atom would block the Rydberg excitation of all the surrounding atoms for $V(R)\geq\delta_{\rm EIT}$, where $\delta_{\rm EIT}=|\Omega_{c}|^2/\gamma_{21}$ is the linewidth of EIT transmission window. Therefore, the blockade sphere~\cite{Petrosyan} has a radius of
$R_{b}=(C_{6}/\delta_{\rm EIT})^{1/6}\approx5.45\,{\rm\mu m}$ for $\Omega_{c}=2\pi\times4\,{\rm MHz}$, and thus has a volume of $V_{b}=(4/3)\pi R_{b}^3\approx678.66\,{\rm\mu m}^{3}$. Comparing this to the average interatomic separation
obtained by ${\bar R}=(5/9)N_{a}^{-1/3}\approx2.42\,{\rm\mu m}$ for $N_{a}=1.2\times10^{-10}\,{\rm cm}^{-3}$~\cite{Pritchard}, we see that the blockade effect can be obviously observed, and the number of atoms inside the blockade radius can be evaluated by $N_b=N_{a}V_{b}\approx8.2$, as shown in Fig.~\ref{Fig1}(c). The system can be divided into many blockade
spheres [represented by the spheres with the boundary indicated by yellow dashed line in
Fig.~\ref{Fig1}(c)] and each blockade sphere contains only one Rydberg atom [represented by the small
yellow sphere in Fig.~\ref{Fig1}(c)].

The Heisenberg equation of motion for ${\hat S}_{\alpha\beta}({\bf r},t)$ is given by $[i\partial/\partial t +(\omega_{\alpha}-\omega_{\beta}+\Delta_{\alpha}-\Delta_{\beta})]{\hat S}_{\alpha\beta}=(1/\hbar)[{\hat S}_{\alpha\beta},
{\hat H}_H]$. Based on this, we can obtain the equation of the 1-body correlators (or called 1-body density matrix elements) $\rho_{\alpha\beta}({\bf r},t)\equiv\langle {\hat S}_{\alpha\beta}({\bf r},t)\rangle$~\cite{note1}
\begin{subequations}
\begin{align}
&i\frac{\partial}{\partial t}\rho_{11}-i\Gamma_{12}\rho_{22}-\Omega_{p}\rho_{12}+\Omega_{p}^{\ast}\rho_{21}=0,\\
&i\left(\frac{\partial}{\partial t}+\Gamma_{12}\right)\rho_{22}-i\Gamma_{23}\rho_{33}+\Omega_{p}\rho_{12}-\Omega_{p}^{\ast}\rho_{21}
-\Omega_{c}\rho_{23}+\Omega_{c}^{\ast}\rho_{32}=0,\\
&i\left(\frac{\partial}{\partial t}+\Gamma_{23}\right)\rho_{33}+\Omega_{c}\rho_{23}-\Omega_{c}^{\ast}\rho_{32}=0,\\
&\left(i\frac{\partial}{\partial t}+d_{21}\right)\rho_{21}+\Omega_{p}(\rho_{11}-\rho_{22})+\Omega_{c}^{\ast}\rho_{31}=0,\\
&\left(i\frac{\partial}{\partial t}+d_{31}\right)\rho_{31}-\Omega_{p}\rho_{32}+\Omega_{c}\rho_{21}-N_{a}\int_{{\bf r}^{\prime}\neq{\bf r}}{d^3{\bf r}^{\prime}V({\bf r}^{\prime}-{\bf r})\rho_{33,31}\left({\bf r}^{\prime},{\bf r},t\right)}=0,\\
&\left(i\frac{\partial }{\partial t}+d_{32}\right)\rho_{32}-\Omega_{p}^{\ast}\rho_{31}+\Omega_{c}\left(\rho_{22}-\rho_{33}\right)-N_{a}\int_{{\bf r}^{\prime}\neq{\bf r}}d^3{\bf r}^\prime V\left({\bf r}^\prime-{\bf r}\right)\rho_{33,32}({\bf r}^{\prime},{\bf r},t)=0,
\end{align}
\label{Heisenberg Eqs}
\end{subequations}
where $d_{\alpha\beta}=\Delta_{\alpha}-\Delta_{\beta}+i\gamma_{\alpha\beta}\,(\alpha,\beta=1,2,3;\alpha\neq\beta)$, $\gamma_{\alpha\beta}=(\Gamma_{\alpha}+\Gamma_{\beta})/2+\gamma_{\alpha\beta}^{\rm dep}$ with $\Gamma_{\beta}=\sum_{\alpha<\beta}\Gamma_{\alpha\beta}$. Here $\Gamma_{\alpha\beta}$ denotes the spontaneous emission decay rate from the state $|\beta\rangle$ to the state $|\alpha\rangle$, and $\gamma_{\alpha\beta}^{\rm dep}$ denotes the dephasing
(including those from atomic motion and the interaction between ground-state and Rydberg-sate atoms)
rate between $|\alpha\rangle$ and $|\beta\rangle$.

From Eq.~(\ref{Heisenberg Eqs}), we see that for solving the equations of motion of the 1-body correlators, we need to know the 2-body correlators (2-body density matrix elements)  $\rho_{33,3\alpha}({\bf r}',{\bf r},t)\equiv \langle{\hat S}_{33}({\bf r}',t){\hat S}_{3\alpha}({\bf r},t)\rangle$ ($\alpha=1,2$). It is easy to show that for solving the equations of motion of the 2-body correlators, we need to know 3-body correlators, defined by $\rho_{\alpha\beta,\mu\nu,
\zeta\eta}({\bf r}^{\prime\prime},{\bf r}^{\prime},t)\equiv \langle{\hat S}_{\alpha\beta}({\bf r}^{\prime\prime},t){\hat S}_{\mu\nu}({\bf r}',t){\hat S}_{\zeta\eta}({\bf r},t)\rangle$,
etc. As a result, we obtain an infinite hierarchy of equations of motion for the correlators of 1-body, 2-body, 3-body, and so on.

\subsection{Transient response of the Rydberg-EIT in many-atom model}~\label{sec3b}
The equations of motion of the 1-body correlators are given in Eq.~(\ref{Heisenberg Eqs}). Equations of motion of 2-body correlators
are not listed here since there are 27 independent equations and each of them is long. In fact, these equations have almost the same forms as those of the two-atom density matrix elements derived in the two-atom model [see Eq.~(\ref{Two-atom Eqs}) of {\color{blue}Appendix~\ref{AppAa}}], but with additional 3-body correlators and corresponding spatial integrals [related to vdW potential $\hbar V({\bf r}'-{\bf r})$] involved. Because these equations are nonlinearly coupled with each other, it is difficult to solve them by using conventional techniques. Fortunately, since in our consideration the probe-field intensity is relatively small and hence we can employ the method of reduction perturbation, widely applied in nonlinear oscillation and wave theory~\cite{Jeffery}, to solve them. Because our calculation is exact to third order (i.e. up to $\Omega_p^3$), the equations of motion for the $n$-body correlators ($n\ge 3$) are not needed.
In principle, one can go to higher orders of $\Omega_p$, valid for large probe field~\cite{BaiOE}, this will, however, involve a large amount of calculations.

%%%%%%%%%%%%%%%%%%%%%%%%%%%%%%%%%%%%%%%%%%%%%%%%%%%%%%%%%%%%%
\subsubsection{Solutions of 1-body and 2-body correlators using a method of multiple-scales}~\label{sec3b1}
By inspection on the order of magnitude in the equations of the 1-body correlators $\rho_{\alpha\beta}\equiv\langle\hat{S}_{\alpha\beta}\rangle$ and the
2-body correlators $\rho_{\alpha\beta,\mu\nu}\equiv\langle{\hat S}_{\alpha\beta}{\hat S}_{\mu\nu}\rangle$, we make the following expansions:
$\Omega_{p}=\epsilon\Omega_{p}^{(1)}$,
$\rho_{\alpha1}=\sum_{m=0}\epsilon^{2m+1}\rho_{\alpha1}^{(2m+1)}$,
$\rho_{\alpha\beta}=\sum_{m=1}\epsilon^{2m}\rho_{\alpha\beta}^{(2m)}$,
$\rho_{11}=1+\sum_{m=1}\epsilon^{2m}\rho_{11}^{(2m)}$,
$\rho_{\alpha1,\beta1}=\sum_{m=1}\epsilon^{2m}\rho_{\alpha1,\beta1}^{(2m)}$, $\rho_{\alpha1,1\beta}=\sum_{m=1}\epsilon^{2m}\rho_{\alpha1,1\beta}^{(2m)}$, $\rho_{\alpha\beta,\mu1}=\sum_{m=1}\epsilon^{2m+1}\rho_{\alpha\beta,\mu1}^{(2m+1)}$, and $\rho_{\alpha\beta,\mu\nu}=\sum_{m=2}\epsilon^{2m}\rho_{\alpha\beta,\mu\nu}^{(2m)}\,(\alpha,\beta,\mu,\nu=2,3)$. Here $\epsilon$ is a small expansion parameter, introduced for characterizing the magnitude of the amplitude of the probe-field Rabi frequency.

To obtain divergence-free solutions for the 1-body and 2-body correlators, all the quantities on the right hand side of the expansions given above are considered as functions of the fast time variable $t_{0}=t$ and the slow time variable $t_2=\epsilon^2 t$~\cite{Newell,Jeffery}. Then we obtain a set of linear but inhomogeneous differential equations for each of the equations of the 1- and 2-body correlators, which can be solved analytically order by order up to third-order approximation.

%%%%%%%%%%%%%%%%%%%%%%%%%%%%%%%%%%%%%%%%%%%%%%%%%%%%%%%%%%%%%%%%%%%%%%%%
At the first [i.e. $O(\epsilon)$] order, only the equations for 1-body correlators are to be solved. By using the
initial condition $\rho_{21}^{(1)}(0)=-\Omega_{p}^{(1)}/d_{21}$, $\rho_{31}^{(1)}(0)=0$, we obtain
the solution for $\rho_{\alpha1}^{(1)}$, which has a damped fast oscillation (as a function of $t_0$) modulated by two envelopes $f_1^{(1)}$ and $f_2^{(1)}$ (as a function of $t_2$) [see Eq.~(\ref{Sols rho1_1}) of {\color{blue}Appendix~\ref{AppB}}].
%%%%%%%%%%%
At the second [i.e. $O(\epsilon^2)$] order, we obtain the lowest-order solution of the 2-body correlators with the given set of initial conditions is $\rho_{21,21}^{(2)}(0)=(\Omega_{p}^{(1)}/d_{21})^2$, $\rho_{21,12}^{(2)}(0)=|\Omega_{p}^{(1)}/d_{21}|^2$ and other $\rho_{\alpha1,\beta1}^{(2)}(0)=\rho_{\alpha1,1\beta}^{(2)}(0)=0$.
The second-order solution for the 1-body correlators $\rho_{\alpha\beta}^{(2)}$ can also be gained simultaneously with the set of initial conditions $\rho_{22}^{(2)}(0)=2\gamma_{21}|\Omega_{p}^{(1)}|^2/(\Gamma_{12}|d_{21}|^2)$ and other $\rho_{\alpha\beta}^{(2)}(0)=0$.
%%%%%%%%%%%%%%%%%%%%%%%%
With these results, we proceed to the third [i.e. $O(\epsilon^3)$] order approximation. Solutions of $\rho_{\alpha\beta,\mu1}^{(3)}$ and $\rho_{\alpha1}^{(3)}$ at this order are to be obtained. A solvability condition (i.e. to cancel the secular term appeared in the third-order equation) is used to get the envelopes $f_1^{(1)}$ and $f_2^{(1)}$ appeared in the first-order solution. Steps for obtaining the second-order and third-order approximated solutions for the equations of the 1-body and 2-body correlators by using the method of multiple-scales have been described in detail in {\color{blue}Appendix~\ref{AppB}}.
%

%%%%%%%%%%%%%%%%%%%%%%%%%%%%%%%%%%%%%%%%%%%%%%%%%%%%%%%
%\vspace{3mm}
\subsubsection{Transient response of the Rydberg-EIT in the many-atom model and a comparison with the two-atom model}~\label{sec3b2}
Combining the solutions gained from the first- to the third-order approximations described above, after returning to the original variables we obtain the transient optical response function of the Rydberg-EIT in the many-atom model, given by
\begin{equation}
\rho_{21}(t)\approx a_{21}^{(1)}(t)\Omega_{p}+\left[N_{a}\int d^3{\bf r}'V({\bf r}'-{\bf r})a_{21}^{(3),\rm RR}({\bf r}'-{\bf r},t)+a_{21}^{(3),\rm LA}(t)\right]|\Omega_{p}|^2\Omega_{p}.
\label{rho21}
\end{equation}
Here the first (second) term on the right hand side is linear (nonlinear) optical response of the system. The nonlinear response includes two parts. One is a nonlocal nonlinear response, described by the $N_{a}\int d^3{\bf r}'V({\bf r}'-{\bf r})a_{21}^{(3),\rm RR}({\bf r}'-{\bf r},t)\,|\Omega_{p}|^2\Omega_{p}$, which is contributed from the Rydberg interaction; Another one is a local nonlinear response, described by the $a_{21}^{(3),\rm LA}(t)\,|\Omega_{p}|^2\Omega_{p}$, which is contributed from the photon-atom interaction.
For detailed expressions of $a_{21}^{(1)}$, $a_{21}^{(3),\rm RR}$, and $a_{21}^{(3),\rm LA}$, see Eq.~(\ref{Sols rho1_1}a), Eq.~(\ref{rho1 3rd Sols}a), and Eq.~(\ref{rho1 3rd Sols}b) of {\color{blue}Appendix~\ref{AppB}}. Note that the local nonlinear response is much smaller than the nonlocal one, and it towards zero if the two-photon detuning $\Delta_3=0$.

Shown in Fig.~\ref{Fig2}(b) is the normalized absorption ${\rm Im}(\rho_{21})$ as a function of $t$ for the many-atom model by taking $\Omega_{p}=0.05\Gamma_{12}$. In the figure, the green dashed-dotted line is for the case $\Omega_{c}=2\pi\times4\,{\rm MHz}$ with a high atomic density ($N_{a}=1.2\times10^{10}\,{\rm cm}^{-3}$) and hence significant Rydberg interaction; the brown dotted line is for the case $\Omega_{c}=2\pi\times4\,{\rm MHz}$ with a low atomic density ($N_{a}=1\times10^{8}\,{\rm cm}^{-3}$) and hence negligible Rydberg interaction.
Results for a large control field, i.e. $\Omega_{c}=2\pi\times8\,{\rm MHz}$, are also shown, with the blue solid and red dashed lines being for the presence and absence of the Rydberg interaction, respectively.

From the figure, we see that:
(i)~Similar to the numerical result obtained from the two-atom model, both the absorption curves of the EIT with and without the Rydberg interaction display a damped oscillation before reaching to a small steady-state value as the control field is switched on.
(ii)~The oscillation amplitude for the case with the Rydberg interaction is smaller (the green dashed-dotted line) compared with the case with no Rydberg interaction (the brown dotted line), the same as that of the numerical result obtained from the two-atom model. However, the decrease of the oscillation amplitude by the Rydberg interaction here is smaller than that in the two-atom model because we have selected a smaller probe field ($\Omega_{p}=0.05\Gamma_{12}$) in order to make the perturbation calculation be valid. We speculate that the oscillation amplitude will increase is $\Omega_{p}$ is taken a larger value.

Shown in Fig.~\ref{Fig3}(d) and Fig.~\ref{Fig3}(e) are numerical results of normalized absorption
spectrum Im$(\rho_{21})$ as a function of probe-field detuning $\Delta$ at $t=0.14\,\rm\mu s$ and $t = 1.2\,\rm\mu s$ for $\Omega_{c}=2\pi\times4\,$MHz and $\Omega_{p}=0.03\Gamma_{12}$, respectively, where the blue solid line is the EIT spectrum with a high atomic density $N_{a}=1.2\times10^{10}\,{\rm cm}^{-3}$ (significant Rydberg interaction), while the red dashed line is the EIT spectrum with a low atomic density $N_{a}=1\times10^{8}\,{\rm cm}^{-3}$ (negligible Rydberg interaction). From the figure, we see that:
(i)~Similar to the result obtained from the two-atom model, the original single-peak absorption spectrum at $t=0$ [Fig.~\ref{Fig3}(a)] evolves into a structure with two peaks (i.e. a EIT transparency window is opened near $\Delta=0$) after the control field is switched on, and the final steady-state in the EIT with the Rydberg interaction is stronger compared with the EIT with no Rydberg interaction.
(ii)~Different from the symmetric two-peak structure obtained with the two-atom model [red dashed lines in Fig.~\ref{Fig3}(d) and Fig.~\ref{Fig3}(e)], the Rydberg-EIT spectrum calculated with the many-atom model always displays asymmetric profiles [blue solid lines in Fig.~\ref{Fig3}(d) and Fig.~\ref{Fig3}(e)] during the time evolution. This is because the Rydberg interaction may give rise to slight deviation of the two-photon resonance, contributed from the (nonlocal) integration that involve all the surrounding atoms, as indicated at the expression of the response function (\ref{rho21}).

\begin{figure}
\includegraphics[width=0.6\columnwidth]{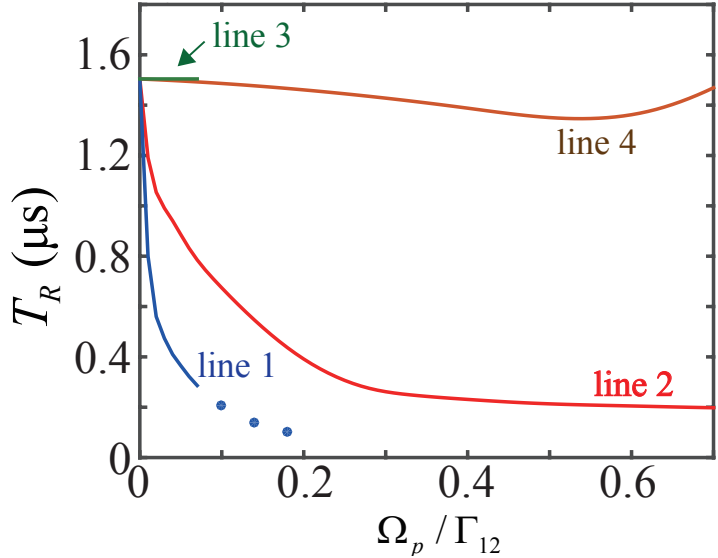}
\caption{\footnotesize (Color online) Response time $T_R$ of the EIT as a function of the probe-field Rabi frequency $\Omega_{p}$. Line 1 (Line 2): $T_R$ for the EIT with the Rydberg interaction obtained in the many-atom (two-atom) model. Line 3 (Line 4): $T_R$ for the EIT with no Rydberg interaction obtained in the many-atom (two-atom) model. Three blue points along the decreasing direction of line 1 indicate the tendency of the response time of the EIT with the Rydberg interaction as $\Omega_{p}$ is increased, which means that $T_R$ will be decreased further for larger probe field.
}
\label{Fig5}
\end{figure}

In order to make a comparison between the results obtained by the many-atom model here and by the two-atom model in the last section, in Fig.~\ref{Fig5} we show
the response time $T_R$ of the EIT as a function of the probe-field Rabi frequency $\Omega_{p}$. Line 1 and line 2 are for the case with the Rydberg interaction, obtained in the many-atom model and the two-atom model, respectively; line 3 and line 4 are for the case with no Rydberg interaction obtained in the many-atom and two-atom models, respectively. When plotting the figure, parameters for the two-atom model are $\Omega_{c}=2\pi\times4\,{\rm MHz}$, $V_{AB}=1.0\,{\rm GHz}$ ($r_{AB}\approx3.10\,{\rm\mu m}$) for the EIT with Rydberg interaction, $V_{AB}=0$ for the EIT without Rydberg interaction. Parameters for the many-atom model are given by $\Omega_{c}=2\pi\times4\,{\rm MHz}$, $N_{a}=1.2\times10^{10}\,{\rm cm}^{-3}$ for the EIT with the Rydberg interaction (adopted from the experiment~\cite{Pritchard}), $N_{a}=1\times10^{8}\,{\rm cm}^{-3}$ for the EIT with negligible Rydberg interaction.

From Fig.~\ref{Fig5} we can arrive the following conclusions:
(1).~The response time of the EIT with the Rydberg interaction is much faster than that of the EIT without the Rydberg interaction.
(2).~For a given probe-field Rabi frequency $\Omega_{p}$, the response time of the Rydberg-EIT (line 1) obtained by the many-atom model is faster than that obtained by the two-atom model (line 2).
(3).~As the probe-field Rabi frequency $\Omega_{p}$ is increased, the response time of the EIT with the Rydberg interaction (lines 1 and 2) is reduced rapidly. However, the response time of the EIT without the Rydberg interaction displays no obvious tendency of reduction when $\Omega_{p}$ increases (lines 3 and 4).

The physical reason for the fast-responding property of the Rydberg-EIT in the many-atom system is mainly due to the Rydberg blockade effect.
Due to this effect, in each blockade sphere only one atom is excited to the Rydberg state $|3\rangle$, other atoms can only be excited to the state $|2\rangle$. Thus in the Rydberg-EIT system, most atoms behave practically like two-level ones, and the system has a larger relaxation rate compared with
the EIT system without the Rydberg interaction. As a result, the dissipation of the system is enhanced (with relaxation rate scaled with $\approx N_b$), giving rise to a decreased response time for the Rydberg-EIT system.

Comparing with the two-atom system, in the many-atom system the Rydberg blockade effect is enhanced much, and hence the response speed of the EIT is faster than that of the two-atom one.
Note that the perturbation calculation presented above, though attained under a weak probe-field approximation ($N_a$ is fixed), can be in principle extended to high orders when $\Omega_p$ (or $N_a$) becomes larger. One expects that the result on the optical response of the Rydberg-EIT given above can be extended to the case of large probe-field intensity. One can make a prediction on the variation tendency of $T_R$ when $\Omega_{p}$ becomes large. Three blue points along the decreasing direction of line 1 indicate the tendency of $T_R$ of the EIT with the Rydberg interaction as $\Omega_{p}$ grows, which means that the EIT response time can be decreased further as the probe field is increased.

\section{Discussion and Summary}~\label{sec4}

We noticed that the transient many-body dynamics of Rydberg atoms has attracted much attention in recent years, including, e.g., coherent Rydberg excitations~\cite{Robicheaux,Olmos}, collectively enhanced Rabi oscillations~\cite{Stanojevic,Dudin}, and suppression of multiple Rydberg excitations~\cite{Johnson,Reetz}, etc. However, our work is very different from Refs.~\cite{Robicheaux,Olmos,Stanojevic,Dudin,Johnson,Reetz}. First, the transient dynamics considered in Refs.~\cite{Robicheaux,Olmos,Stanojevic,Dudin,Johnson,Reetz}
is outside of EIT regime, whereas what we considered here is inside an EIT regime.
Second, the atomic model used in Refs.~\cite{Robicheaux,Olmos,Stanojevic,Dudin,Johnson,Reetz} is either a two-level or a three-level one with a very large one-photon detuning, whereas in our model no constraint on the one-photon detuning is used. Third, light fields used in Refs.~\cite{Robicheaux,Olmos,Stanojevic,Dudin,Johnson,Reetz}
must be assumed to be strong enough so that they can be taken to be undepleted during Rydberg excitations and transient response processes, whereas in our work the probe field used is weak and thus the optical
susceptibilities of the system during the Rydberg excitation and the transient response process can be obtained both analytically and numerically.

In conclusion, we have studied the transient optical response property of the EIT in a cold Rydberg atomic gas with the Rydberg interaction. We have demonstrated that both the transient behavior and the steady-state EIT spectrum of the system depend on the Rydberg interaction strongly. In particular, the response speed of the Rydberg-EIT may be five-times faster than the conventional EIT without the Rydberg interaction, and can be increased further by increasing the probe-field intensity. For comparison, two different models (i.e. two-atom model and many-atom model) are solved. The results reveal that Rydberg blockade effect plays a significant role for increasing the response speed of the Rydberg-EIT. The fast-responding Rydberg-EIT by using the strong, tunable Rydberg interaction found here is useful not only for a deep understanding of the non-equilibrium many-body dynamics of Rydberg atoms, but also for practical applications in quantum information processing (including all-optical switching and transistors, quantum phase gates, etc.) based on Rydberg atoms.

\acknowledgments This work was supported by the NSF-China under Grant No. 11174080, China Postdoctoral Science Foundation funded project under Grant No. 2017M620140, and by the 111 Project under Grant No. B12024.

%%%%%%%%%%%%%%%%%%%%%%%%%%%%%%%%%%%%%%%%%
\appendix
\section{Two-atom model}\label{AppA}
\subsection{Equations of motion for two-atom density matrix elements}\label{AppAa}
The explicit form of the master equation Eq.~(\ref{Schrodinger Eqs}) in the main text reads
\begin{subequations}
\begin{align}
% sgmsgm{11,11}
& i\frac{\partial}{\partial t}\rho_{11,11}-2i\Gamma_{12}\rho_{11,22}-2\Omega_{p}\rho_{12,11}
+2\Omega_{p}^{\ast}\rho_{21,11}=0,\nonumber\\
% sgmsgm{22,11}
& \nonumber i\left(\frac{\partial}{\partial t}+\Gamma_{12}\right)\rho_{22,11}-i\Gamma_{12}\rho_{22,22}-i\Gamma_{23}\rho_{11,33}
+\Omega_{p}\rho_{12,11}-\Omega_{p}\rho_{22,12}\\
& \ \ \ \ -\Omega_{p}^{\ast}\rho_{21,11}+\Omega_{p}^{\ast}\rho_{22,21}-\Omega_{c}\rho_{23,11}
+\Omega_{c}^{\ast}\rho_{32,11}=0,\nonumber\\
% sgmsgm{33,11}
& \nonumber i\left(\frac{\partial}{\partial t}+\Gamma_{23}\right)\rho_{33,11}-i\Gamma_{12}\rho_{33,22}+\Omega_{c}\rho_{23,11}-\Omega_{c}^{\ast}\rho_{32,11}
-\Omega_{p}\rho_{33,12}+\Omega_{p}^{\ast}\rho_{33,21}=0,\nonumber\\
% sgmsgm{21,11}
& \nonumber \left(i\frac{\partial}{\partial t}+d_{21}\right)\rho_{21,11}-i\Gamma_{12}\rho_{21,22}+\Omega_{p}\left(\rho_{11,11}
-\rho_{22,11}\right)\\
& \ \ \ \ +\Omega_{c}^{\ast}\rho_{31,11}-\Omega_{p}\rho_{21,12}
+\Omega_{p}^{\ast}\rho_{21,21}=0,\nonumber\\
% sgmsgm{31,11}
& \nonumber \left(i\frac{\partial}{\partial t}+d_{31}\right)\rho_{31,11}-i\Gamma_{12}\rho_{31,22}+\Omega_{c}\rho_{21,11}-\Omega_{p}\rho_{32,11}
-\Omega_{p}\rho_{31,12}+\Omega_{p}^{\ast}\rho_{31,21}=0,\nonumber\\
% sgmsgm{32,11}
& \nonumber \left(i\frac{\partial}{\partial t}+d_{32}\right)\rho_{32,11}-i\Gamma_{12}\rho_{32,22}+\Omega_{c}\left(\rho_{22,11}
-\rho_{33,11}\right)\\
& \ \ \ \ -\Omega_{p}\rho_{32,12}+\Omega_{p}^{\ast}\rho_{32,21}-\Omega_{p}^{\ast}\rho_{31,11}=0,\nonumber\\
% sgmsgm{21,21}
& \left(i\frac{\partial }{\partial t}+2d_{21}\right)\rho_{21,21}-2\Omega_{p}\left(\rho_{22,21}-\rho_{11,21}\right)+2\Omega_{c}^\ast\rho_{31,21}=0,\nonumber\\
% sgmsgm{21,12}
& \nonumber \left(i\frac{\partial }{\partial t}+d_{21}+d_{12}\right)\rho_{21,12}+\Omega_{p}\left(\rho_{11,12}-\rho_{22,12}\right)\\
& \ \ \ \ +\Omega_{p}^{\ast}\left(\rho_{22,21}-\rho_{11,21}\right)-\Omega_{c}\rho_{21,13}+\Omega_{c}^\ast\rho_{31,12}=0,\nonumber\\
% sgmsgm{21,31}
& \nonumber \left(i\frac{\partial }{\partial t}+d_{21}+d_{31}\right)\rho_{21,31}+\Omega_{c}\rho_{21,21}+\Omega_{c}^{\ast}\rho_{31,31}
-\Omega_{p}(\rho_{21,32}+2\rho_{22,31}+\rho_{33,31}-\rho_{31})=0, \nonumber\\
% sgmsgm{21,13}
& \nonumber \left(i\frac{\partial }{\partial t}+d_{21}+d_{13}\right)\rho_{21,13}+\Omega_{c}^\ast(\rho_{31,13}-\rho_{21,12})
-\Omega_{p}\left(2\rho_{22,13}+\rho_{33,13}-\rho_{13}\right)+\Omega_{p}^\ast\rho_{21,23}=0, \nonumber\\
% sgmsgm{31,31}
&\left(i\frac{\partial }{\partial t}+2d_{31}-V_{12}\right)\rho_{31,31}-2\Omega_{p}\rho_{32,31}+2\Omega_{c}\rho_{21,31}=0, \nonumber\\
% sgmsgm{31,13}
&\left(i\frac{\partial }{\partial t}+d_{31}+d_{13}\right)\rho_{31,13}-\Omega_{p}\rho_{32,13}+\Omega_{p}^{\ast}\rho_{31,23}+\Omega_{c}\rho_{21,13}
-\Omega_{c}^{\ast}\rho_{31,12}=0,\nonumber\\
% \rho_{22,21}
& \nonumber \left(i\frac{\partial }{\partial t}+i\Gamma_{12}+d_{21}\right)\rho_{22,21}-i\Gamma_{23}\rho_{33,21}-\Omega_{c}\rho_{23,21}
+\Omega_{c}^{\ast}(\rho_{32,21}+\rho_{22,31}) \\
& \ \ \ \ -\Omega_{p}^{\ast}\rho_{21,21}-\Omega_{p}\left(\rho_{22,22}-\rho_{22,11}-\rho_{12,21}\right)=0, \nonumber\\
% \rho_{22,31}
& \nonumber \left(i\frac{\partial }{\partial t}+i\Gamma_{12}+d_{31}\right)\rho_{22,31}-i\Gamma_{23}\rho_{33,31}+\Omega_{c}(\rho_{22,21}-\rho_{23,31})+\Omega_{c}^{\ast}\rho_{32,31}\\
& \ \ \ \ +\Omega_{p}(\rho_{12,31}-\rho_{22,32})-\Omega_{p}^{\ast}\rho_{21,31}=0, \nonumber\\
% \rho_{33,21}
& \nonumber \left(i\frac{\partial }{\partial t}+i\Gamma_{23}+d_{21}\right)\rho_{33,21}+\Omega_{c}\rho_{23,21}+\Omega_{c}^{\ast}\left(\rho_{33,31}
-\rho_{32,21}\right)-\Omega_{p}\left(\rho_{33,22}-\rho_{33,11}\right)=0, \nonumber\\
% \rho_{33,31}
& \nonumber \left(i\frac{\partial}{\partial t}+i\Gamma_{23}+d_{31}-V_{12}\right)\rho_{33,31}-\Omega_{c}^{\ast}\rho_{32,31}
+{\Omega}_{c}(\rho_{33,21}+\rho_{23,31})-\Omega_{p}\rho_{33,32}=0, \nonumber\\
% \rho_{32,21}
& \nonumber \left(i\frac{\partial }{\partial t}+d_{32}+d_{21}\right)\rho_{32,21}-\Omega_{c}\left(\rho_{33,21}-\rho_{22,21}\right)
+\Omega_c^\ast\rho_{32,31}\\
& \ \ \ \ -\Omega_{p}^\ast\rho_{31,21}-\Omega_{p}\left(\rho_{32,22}-\rho_{32,11}\right)=0, \nonumber\\
% \rho_{32,31}
& \nonumber \left(i\frac{\partial }{\partial t}+d_{32}+d_{31}-V_{12}\right)\rho_{32,31}-\Omega_{p}^{\ast}\rho_{31,31}
-\Omega_{p}\rho_{32,32}-\Omega_{c}\left(\rho_{33,31}-\rho_{22,31}-\rho_{32,21}\right)=0, \nonumber\\
% \rho_{23,21}
& \nonumber \left(i\frac{\partial }{\partial t}+d_{23}+d_{21}\right)\rho_{23,21}-\Omega_{p}\left(\rho_{23,22}-\rho_{23,11}-\rho_{13,21}\right)
+\Omega_{c}^{\ast}\left(\rho_{33,21}-\rho_{22,21}+\rho_{23,31}\right)=0, \nonumber\\
% \rho_{23,31}
& \nonumber \left(i\frac{\partial }{\partial t}+d_{23}+d_{31}\right)\rho_{23,31}+\Omega_{p}\rho_{13,31}+\Omega_{c}^{\ast}\left(\rho_{33,31}-\rho_{22,31}\right)
+\Omega_{c}\rho_{23,21}-\Omega_{p}\rho_{23,32}=0,\nonumber\\
%
%
%% start from the fourth order
% sgmsgm{22,22}
& \nonumber i\left(\frac{\partial}{\partial t}+2\Gamma_{12}\right)\rho_{22,22}-2i\Gamma_{23}\rho_{22,33}+2\Omega_{p}\rho_{22,12}-2\Omega_{p}^{\ast}\rho_{22,21}
+2\Omega_{c}^{\ast}\rho_{32,22}-2\Omega_{c}\rho_{22,23}=0,\nonumber\\
% sgmsgm{22,33}
& \nonumber i\left(\frac{\partial}{\partial t}+\Gamma_{23}+\Gamma_{12}\right)\rho_{22,33}-i\Gamma_{23}\rho_{33,33}
+\Omega_{p}\rho_{12,33}-\Omega_{p}^{\ast}\rho_{21,33}\\
& \ \ \ \ -\Omega_{c}\rho_{23,33}+\Omega_{c}^{\ast}\rho_{32,33}
+\Omega_{c}\rho_{22,23}-\Omega_{c}^{\ast}\rho_{22,32}=0,\nonumber\\
% sgmsgm{33,33}
& i\left(\frac{\partial}{\partial t}+2\Gamma_{23}\right)\rho_{33,33}+2\Omega_{c}\rho_{23,33}-2\Omega_{c}^{\ast}\rho_{32,33}=0,\nonumber\\
% sgmsgm{22,32}
& \nonumber \left(i\frac{\partial}{\partial t}+i\Gamma_{12}+d_{32}\right)\rho_{22,32}-i\Gamma_{23}\rho_{33,32}+\Omega_{p}\rho_{12,32}-\Omega_{p}^{\ast}\rho_{21,32}\\
& \ \ \ \ -\Omega_{c}\rho_{23,32}+\Omega_{c}^{\ast}\rho_{32,32}+\Omega_{c}\left(\rho_{22,22}-\rho_{22,33}\right)
-\Omega_{p}^{\ast}\rho_{22,31}=0,\nonumber\\
% sgmsgm{33,32}
& \nonumber \left(i\frac{\partial}{\partial t}+i\Gamma_{23}+d_{32}-V_{12}\right)\rho_{33,32}+\Omega_{c}\rho_{23,32}-\Omega_{c}^{\ast}\rho_{32,32}
-\Omega_{p}^{\ast}\rho_{33,31}+\Omega_{c}\left(\rho_{33,22}-\rho_{33,33}\right)=0,\nonumber\\
% sgmsgm{32,32}
& \left(i\frac{\partial}{\partial t}+2d_{32}-V_{12}\right)\rho_{32,32}+2\Omega_{c}\left(\rho_{22,32}-\rho_{33,32}\right)
-2\Omega_{p}^{\ast}\rho_{31,32}=0,\nonumber\\
% sgmsgm{32,23}
& \nonumber \left(i\frac{\partial}{\partial t}+d_{32}+d_{23}\right)\rho_{32,23}
+\Omega_{c}\left(\rho_{22,23}-\rho_{33,23}\right)
-\Omega_{c}^{\ast}\left(\rho_{22,32}-\rho_{33,32}\right)\\
& \ \ \ \ +\Omega_{p}\rho_{32,13}-\Omega_{p}^{\ast}\rho_{31,23}=0,\nonumber
\end{align}
\label{Two-atom Eqs}
\end{subequations}
where $d_{\alpha\beta}=\Delta_{\alpha}-\Delta_{\beta}+i\gamma_{\alpha\beta}$ with $\gamma_{\alpha\beta}=(\Gamma_{\alpha}+\Gamma_{\beta})/2+\gamma_{\alpha\beta}^{\rm dep}$ and $\Gamma_{\beta}=\sum_{\alpha<\beta}\Gamma_{\alpha\beta}$. Here $\Gamma_{\alpha\beta}$ denotes the spontaneous emission decay rate from the state $|\beta\rangle$ to the state $\alpha\rangle$, and $\gamma_{\alpha\beta}^{\rm dep}$ represents the dephasing rate, resulted, e.g.,  from the atomic motion and the interaction between the atoms in the ground state and the atoms in the Rydberg state.
With the initial condition $\rho_{\alpha\beta,\mu\nu}(0)=\rho_{\alpha\beta}^{A}(0)\rho_{\mu\nu}^{B}(0)$ (for their expressions, see Sec.~\ref{sec2b} in the main text), the above motion equations can be solved numerically by using Runge-Kutta method.

%%%%%%%%%%%%%%%%%%%%%%%%%%%%%%%%%%%%%%%%%%%%%%%%%%%%%%%%%%%%%%%%%%%%%%%%%%%%%%%%%%%%%%%%
\subsection{Dispersion property of the system}\label{AppAb}
\begin{figure}
\includegraphics[width=0.8\columnwidth]{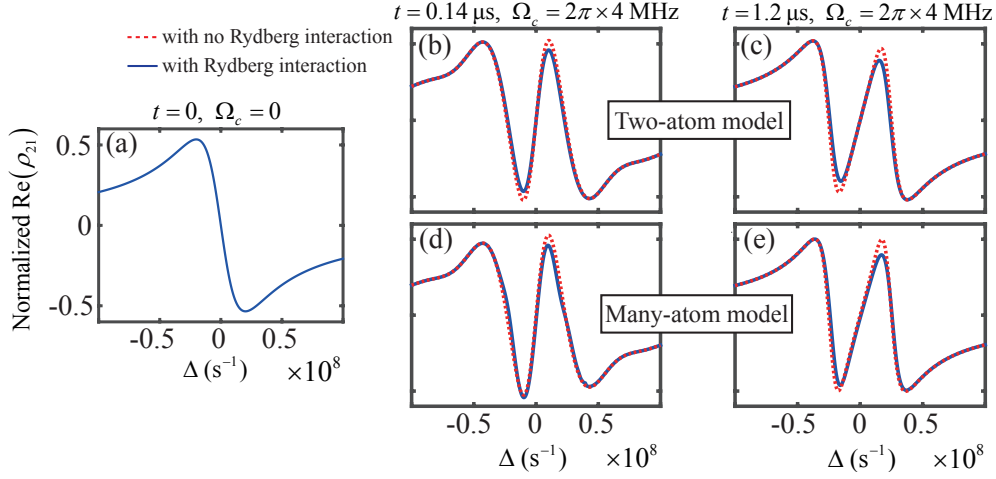}
\caption{\footnotesize  (Color online) Transient response behavior of the Rydberg-EIT
as a function of the probe-field detuning $\Delta\,(\equiv\Delta_{2}=\Delta_{3})$.
(a) Normalized dispersion spectrum ${\rm Re}(\rho_{21})$  for $t=0$, where $\Omega_c=0$.
(b)~${\rm Re}(\rho_{21})$ at $t=0.14\,{\rm\mu s}$ for $\Omega_c=2\pi\times 4$\,MHz.
(c)~${\rm Re}(\rho_{21})$ at $t=1.2\,{\rm\mu s}$ for $\Omega_c=2\pi\times 4$\,MHz.
Both (b) and (c) are obtained from the two-atom model with $\Omega_{p}=0.2\Gamma_{12}$, where the blue solid line is the EIT spectrum with the Rydberg interaction $V_{AB}=1\,{\rm GHz}$ ($r_{AB}=3.10\,{\rm\mu m}$), while the red dashed line is the EIT spectrum without the Rydberg interaction ($V_{AB}=0$).
(d) and (e) are respectively the same with (a) and (b), but obtained by the many-atom model
with $\Omega_{p}=0.03\Gamma_{12}$, where the blue solid line is the EIT spectrum with a high atomic density $N_{a}=1.2\times10^{10}\,{\rm cm}^{-3}$ (significant Rydberg interaction), while the red dashed line is the EIT spectrum with a low atomic density $N_{a}=1\times10^{8}\,{\rm cm}^{-3}$ (negligible Rydberg interaction)
}
\label{Fig6}
\end{figure}

The dispersion property of the system is described by the real part of the atomic coherence $\rho_{21}$, i.e. Re($\rho_{21})$, as a function of the detuning $\Delta(\equiv \Delta_2=\Delta_3$). Shown in Fig.~\ref{Fig6} are results of normalized dispersion spectrum Re($\rho_{21})$, obtained by using the two-atom model and the many-atom model, respectively.
From the figure we have the following conclusions:
(i)~When the control field is switched on ($t>0$), the dispersion spectrum at $t=0$, which displays
an anomalous dispersion [Fig.~\ref{Fig6}(a)], evolves into the one with a normal dispersion near $\Delta=0$;
(ii)~For both the two-atom model [Fig.~\ref{Fig6}(b) and Fig.~\ref{Fig6}(c)] and the many-atom model [Fig.~\ref{Fig6}(d) and Fig.~\ref{Fig6}(e)], near $\Delta=0$ there is a only small difference of the dispersion behavior between the case with (blue solid lines) and without the Rydberg interaction (red dashed lines). Thus, one can obtain slow group velocity by using the Rydberg-EIT, useful for the slowdown and memory of optical pulses.

\subsection{Definition of the response time for a transient response process}\label{AppAc}
\begin{figure}
\includegraphics[width=0.9\columnwidth]{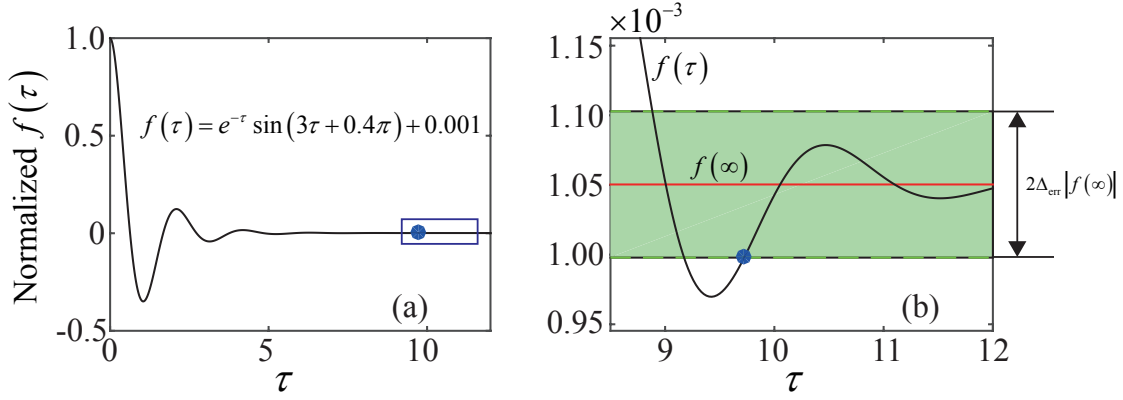}
\caption{\footnotesize  (Color online) Definition of the response time of a transient response process, described by a response function $f(\tau)$, $\tau$ is dimensionless time. (a) Example: $f(\tau)=e^{-\tau}\sin(3\tau+0.4\pi)+0.001$. The blue rectangle indicates that the variation of $f(\tau)$ reaches within the range $2\Delta_{\rm err}$ around the steady-state value $f(\infty)=0.001$. The blue point denotes the response time. (b) Amplification of the blue rectangle shown in (a). The region with green color is the permitted relative error range for defining  response time, marked by the upper boundary $f(\infty)+|f(\infty)|\Delta_{\rm err}$ and the lower boundary $f(\infty)-|f(\infty)|\Delta_{\rm err}$, with $\Delta_{\rm err}=0.05$. The blue point is the response time of the transient response process.}
\label{Fig7}
\end{figure}

To quantitatively determine the response time of a transient response process, one must have a working definition on it. According to engineering control theory (see Refs.~\cite{Ogata,Levine} for detail), the response (or settling) time $T_R$ of a transient response process is usually
defined to be the minimum time after which the temporal change of the response function describing the transient response process always remains within a small error range $2\Delta_{\rm err}$ around the steady-state value of the response. Usually, $\Delta_{\rm err}$ is set to be $0.05$ without loss of generality~\cite{Ogata}. A simple example for the definition of the response time of a response function (denoted by the blue point) is shown in Fig.~\ref{Fig7},
where the normalized response function is $f(\tau)=e^{-\tau}\sin(3\tau+0.4\pi)+0.001$, with $\tau$ the dimensionless time. The blue rectangle in Fig.~\ref{Fig7}(a) means that the variation of $f(\tau)$ has reached into the stage where the variation of $f(\tau)$ is within the range $2\Delta_{\rm err}|f(\infty)|$ around the steady-state value of the response function, i.e. $f(\infty)=0.001$. The blue point indicates the position of the response time. Fig.~\ref{Fig7}(b) is the amplification of the blue rectangle shown in Fig.~\ref{Fig7}(a). The region with green color is the permitted relative error range for determining the response time, marked by the upper boundary $f(\infty)+|f(\infty)|\Delta_{\rm err}$ and the lower boundary $f(\infty)-|f(\infty)|\Delta_{\rm err}$, with $\Delta_{\rm err}=0.05$. Thus the blue point is, by definition, the response time of the transient response process.

%%%%%%%%%%%%%%%%%%%%%%%%%%%%%%%%%%%%%%%%%%%%%
\section{Many-atom model}\label{AppB}
Steps for solving the equations of motion for 1-body and 2-body correlators in the many-atom model are the following:

{\it First-order approximation}. At this order, we need to obtain the 1-body correlators $\rho_{\alpha1}^{(1)}\equiv a_{\alpha1}^{(1)}\Omega_{p}^{(1)}\,(\alpha=2,3)$ only, which satisfy the equation
\begin{equation}
-i\frac{\partial}{\partial t_{0}}\left[
\begin{array}{c}
a_{21}^{(1)} \\
a_{31}^{(1)}
\end{array}
\right]=\left[
\begin{array}{cc}
d_{21} & \Omega_{c}^{\ast} \\
\Omega_{c} & d_{31}
\end{array}
\right]\left[
\begin{array}{c}
a_{21}^{(1)} \\
a_{31}^{(1)}
\end{array}
\right]+\left[
\begin{array}{c}
1 \\
0
\end{array}
\right],
\label{rho1 1st Eqs}
\end{equation}
with the initial condition $a_{21}^{(1)}(0)=-1/d_{21}$, $a_{31}^{(1)}(0)=0$. Here $t_0=t$ is fast time variable. Solution of Eq.~(\ref{rho1 1st Eqs}),
which can be obtained by using constant-variation method~\cite{Jeffery}, reads
\begin{subequations}\label{Sols rho1_1}
\begin{eqnarray}
&& a_{21}^{(1)}=\sum_{m=1}^{2}v_{1m}f_{m}^{(1)}g_{m}^{(1)}e^{i\lambda_{m}t_{0}}+a_{21}^{(1)}(\infty),\\
&& a_{31}^{(1)}=\sum_{m=1}^{2}v_{2m}f_{m}^{(1)}g_{m}^{(1)}e^{i\lambda_{m}t_{0}}+a_{31}^{(1)}(\infty).
\end{eqnarray}
\end{subequations}
Here $a_{21}^{(1)}(\infty)=d_{31}/D$ and $a_{31}^{(1)}(\infty)=-\Omega_{c}/D$ are corresponding steady-state solution~\cite{BaiOE}, with $D=|\Omega_{c}|^2-d_{21}d_{31}$; $g_{m}^{(1)}$ are determined from the initial condition, given by $g_{1}^{(1)}=\{v_{22}[a_{21}^{(1)}(0)-a_{21}^{(1)}(\infty)]-v_{12}[a_{31}^{(1)}(0)-a_{31}^{(1)}(\infty)]\}/(v_{11}v_{22}-v_{21}v_{12})$ and $g_{2}^{(1)}=\{v_{11}[a_{31}^{(1)}(0)-a_{31}^{(1)}(\infty)]-v_{21}[a_{21}^{(1)}(0)-a_{21}^{(1)}(\infty)]\}/(v_{11}v_{22}-v_{21}v_{12})$; $v_{1m}=\Omega_{c}^{\ast}$ and $v_{2m}=\lambda_{m}-d_{21}$, with
\begin{subequations}
\begin{align}
\lambda_{1}=\frac{d_{21}+d_{31}+\sqrt{4|\Omega_{c}|^2+(d_{21}-d_{31})^2}}{2},\\
\lambda_{2}=\frac{d_{21}+d_{31}-\sqrt{4|\Omega_{c}|^2+(d_{21}-d_{31})^2}}{2};
\end{align}
\end{subequations}
$f_{m}^{(1)}$ ($m=1,2$) are slowly-varying envelopes (i.e. functions of the slow time variable $t_2$), yet to be determined in next orders.

{\it Second-order approximation}. We shall obtain the lowest-order solution of the 2-body correlators starts at this order. The first set of equations governing the 2-body correlators $\rho_{\alpha1,\beta1}^{(2)}\equiv a_{\alpha1,\beta1}^{(2)}[\Omega_{p}^{(1)}]^2\,(\alpha,\beta=2,3)$ is given by
\begin{equation}
-i\frac{\partial}{\partial t_{0}}\left[
\begin{array}{c}
a_{21,21}^{(2)} \\
a_{21,31}^{(2)} \\
a_{31,31}^{(2)}
\end{array}
\right]=\left[
\begin{array}{ccc}
2d_{21} & 2\Omega_{c}^{\ast} & 0 \\
\Omega_{c} & d_{21}+d_{31} & \Omega_{c}^{\ast} \\
0 & 2\Omega_{c} & 2d_{31}-V
\end{array}
\right]\left[
\begin{array}{c}
a_{21,21}^{(2)} \\
a_{21,31}^{(2)} \\
a_{31,31}^{(2)}
\end{array}
\right]+\left[
\begin{array}{c}
2a_{21}^{(1)} \\
a_{31}^{(1)} \\
0
\end{array}
\right],
\label{rho2 2nd Eqs}
\end{equation}
with the initial condition $a_{21,21}^{(2)}(0)=1/d_{21}^2$, $a_{21,31}^{(2)}(0)=a_{31,31}^{(2)}(0)=0$.

The second set of equations governing the 2-body correlators $\rho_{\alpha1,1\beta}^{(2)}\equiv a_{\alpha1,1\beta}^{(2)}|\Omega_{p}^{(1)}|^2\,(\alpha,\beta=2,3)$ reads
\begin{equation}
-i\frac{\partial}{\partial t_{0}}\left[
\begin{array}{c}
a_{21,12}^{(2)} \\
a_{31,13}^{(2)} \\
a_{31,12}^{(2)} \\
a_{21,13}^{(2)}
\end{array}
\right]=\left[
\begin{array}{cccc}
2i\gamma_{21} & 0 & \Omega_{c}^{\ast} & -\Omega_{c} \\
0 & 2i\gamma_{31} & -\Omega_{c}^{\ast} & \Omega_{c} \\
\Omega_{c} & -\Omega_{c} & D_{32} & 0 \\
-\Omega_{c}^{\ast} & \Omega_{c}^{\ast} & 0 & D_{23}
\end{array}
\right]\left[
\begin{array}{c}
a_{21,12}^{(2)} \\
a_{31,13}^{(2)} \\
a_{31,12}^{(2)} \\
a_{21,13}^{(2)}
\end{array}
\right]+\left[
\begin{array}{c}
a_{21}^{(1)\ast}-a_{21}^{(1)} \\
0 \\
a_{31}^{(1)\ast} \\
-a_{31}^{(1)}
\end{array}
\right],
\label{rho3 2nd Eqs}
\end{equation}
with $D_{\alpha\beta}=d_{\alpha1}+d_{1\beta}$, where the initial condition is $a_{21,12}^{(2)}(0)=1/|d_{21}|^2$, $a_{21,13}^{(2)}(0)=a_{31,12}^{(2)}(0)=a_{31,13}^{(2)}(0)=0$.

The equation of the 1-body correlators  $\rho_{\alpha\beta}^{(2)}\equiv a_{\alpha\beta}^{(2)}|\Omega_{p}^{(1)}|^2\,(\alpha,\beta=2,3)$ at this order is
\begin{equation}
-i\frac{\partial}{\partial t_{0}}\left[
\begin{array}{c}
a_{22}^{(2)} \\
a_{33}^{(2)} \\
a_{32}^{(2)} \\
a_{23}^{(2)}
\end{array}
\right]=\left[
\begin{array}{cccc}
i\Gamma_{12} & -i\Gamma_{23} & \Omega_{c}^{\ast} & -\Omega_{c} \\
0 & i\Gamma_{23} & -\Omega_{c}^{\ast} & \Omega_{c} \\
\Omega_{c} & -\Omega_{c} & d_{32} & 0 \\
-\Omega_{c}^{\ast} & \Omega_{c}^{\ast} & 0 & d_{23}
\end{array}
\right]\left[
\begin{array}{c}
a_{22}^{(2)} \\
a_{33}^{(2)} \\
a_{32}^{(2)} \\
a_{23}^{(2)}
\end{array}
\right]+\left[
\begin{array}{c}
a_{21}^{(1)\ast}-a_{21}^{(1)} \\
0 \\
a_{31}^{(1)\ast} \\
-a_{31}^{(1)}
\end{array}
\right],
\label{rho4 2nd Eqs}
\end{equation}
with the initial condition $a_{22}^{(2)}(0)=2\gamma_{21}/(\Gamma_{12}|d_{21}|^2)$, $a_{32}^{(2)}(0)=a_{23}^{(2)}(0)=a_{33}^{(2)}(0)=0$. The solution of $\rho_{11}^{(2)}$ is given by $\rho_{11}^{(2)}=-\rho_{22}^{(2)}-\rho_{33}^{(2)}$.

Using Eq.~(\ref{Sols rho1_1}), Eqs.~(\ref{rho2 2nd Eqs})-(\ref{rho4 2nd Eqs}) can be solved by employing the constant-variation method~\cite{Jeffery}. The first term on the right hand side (RHS) of these equations contributes solutions from corresponding homogeneous equations (i.e. in the absence of the second term), while the second term yields inhomogeneous particular solutions.

{\it Third-order approximation}. At this order, equations governing the 2-body correlators $\rho_{\alpha\beta,\mu1}^{(3)}\equiv a_{\alpha\beta,\mu1}^{(3)}|\Omega_{p}^{(1)}|^2\Omega_{p}^{(1)}\,(\alpha,\beta,\mu=2,3)$ are given by
\begin{align}
\nonumber -i\frac{\partial}{\partial t_{0}}\left[
\begin{array}{c}
a_{22,21}^{(3)} \\
a_{22,31}^{(3)} \\
a_{33,21}^{(3)} \\
a_{33,31}^{(3)} \\
a_{32,21}^{(3)} \\
a_{32,31}^{(3)} \\
a_{23,21}^{(3)} \\
a_{23,31}^{(3)}
\end{array}
\right]&=\left[
\begin{array}{cccccccc}
M_{51} & \Omega_{c}^{\ast} & -i\Gamma_{23} & 0 & \Omega_{c}^{\ast} & 0 & -\Omega_{c} & 0 \\
\Omega_{c} & M_{52} & 0 & -i\Gamma_{23} & 0 & \Omega_{c}^{\ast} & 0 & -\Omega_{c} \\
0 & 0 & M_{53} & \Omega_{c}^{\ast} & -\Omega_{c}^{\ast} & 0 & \Omega_{c} & 0 \\
0 & 0 & \Omega_{c} & M_{54} & 0 & -\Omega_{c}^{\ast} & 0 &  \Omega_{c} \\
\Omega_{c} & 0 & -\Omega_{c} & 0 & M_{55} & \Omega_{c}^{\ast} & 0 & 0 \\
0 & \Omega_{c} & 0 & -\Omega_{c} & \Omega_{c} & M_{56} & 0 & 0 \\
-\Omega_{c}^{\ast} & 0 & \Omega_{c}^{\ast} & 0 & 0 & 0 & M_{57} & \Omega_{c}^{\ast} \\
0 & -\Omega_{c}^{\ast} & 0 & \Omega_{c}^{\ast} & 0 & 0 & \Omega_{c} & M_{58}
\end{array}
\right]\left[
\begin{array}{c}
a_{22,21}^{(3)} \\
a_{22,31}^{(3)} \\
a_{33,21}^{(3)} \\
a_{33,31}^{(3)} \\
a_{32,21}^{(3)} \\
a_{32,31}^{(3)} \\
a_{23,21}^{(3)} \\
a_{23,31}^{(3)}
\end{array}
\right] \\
&+\left[
\begin{array}{c}
a_{22}^{(2)}+a_{21,12}^{(2)}-a_{21,21}^{(2)} \\
a_{31,12}^{(2)}-a_{21,31}^{(2)} \\
a_{33}^{(2)} \\
0 \\
a_{32}^{(2)}-a_{21,31}^{(2)} \\
-a_{31,31}^{(2)} \\
a_{23}^{(2)}+a_{21,13}^{(2)} \\
a_{31,13}^{(2)}
\end{array}
\right],
\label{rho5 3th Eqs}
\end{align}
with $M_{51}=i\Gamma_{12}+d_{21}$, $M_{52}=i\Gamma_{12}+d_{31}$, $M_{53}=i\Gamma_{23}+d_{21}$, $M_{54}=i\Gamma_{23}+d_{31}-V$, $M_{55}=d_{32}+d_{21}$, $M_{56}=d_{32}+d_{31}-V$, $M_{57}=d_{23}+d_{21}$, and $M_{58}=d_{23}+d_{31}$. Here the initial condition is given by $a_{22,21}^{(3)}(0)=a_{22}^{(2)}(0)a_{21}^{(1)}(0)$, other $a_{\alpha\beta,\mu1}^{(3)}(0)=0$. With the solutions obtained at the second-order approximation, solutions of these equations can be also obtained analytically.

With the results obtained above, we can proceed to the equations of the 1-body correlators at the third-order approximation, i.e. $\rho_{\alpha1}^{(3)}\equiv a_{\alpha1}^{(3)}|\Omega_{p}^{(1)}|^2\Omega_{p}^{(1)}$, given by
\begin{align}
\nonumber-i\frac{\partial}{\partial t_{0}}\left[
\begin{array}{c}
a_{21}^{(3)} \\
a_{31}^{(3)}
\end{array}
\right]&=\left[
\begin{array}{cc}
d_{21} & \Omega_{c}^{\ast} \\
\Omega_{c} & d_{31}
\end{array}
\right]\left[
\begin{array}{c}
a_{21}^{(3)} \\
a_{31}^{(3)}
\end{array}
\right]+\frac{i}{|\Omega_{p}^{(1)}|^2}\frac{\partial}{\partial t_{2}}\left[
\begin{array}{c}
a_{21}^{(1)} \\
a_{31}^{(1)}
\end{array}
\right]\\
&+\left[
\begin{array}{c}
-2a_{22}^{(2)}-a_{33}^{(2)} \\
-a_{32}^{(2)}-N_{a}\int d^3{\bf r}^{\prime}V\left({\bf r}^{\prime}-{\bf r}\right)a_{33,31}^{(3)}
\end{array}
\right].
\label{rho1 3rd Eqs}
\end{align}
The solution of this equation can be obtained by using the constant-variation method~\cite{Jeffery}, which include the solution of the corresponding homogeneous equation and the particular solution contributed by the inhomogeneous terms (i.e. the second and third terms on the RSH in the equation). Note that the homogeneous equation [i.e. without the second and the third terms on the RHS] has the same eigenvalues and eigenfunctions as those of Eq.~(\ref{rho1 1st Eqs}); the third term on the RHS contributes a similar eigen-oscillation (i.e. resonant drive) to the system, which will results in a secular term in the equation and hence a singularity in its solution. The aim introducing the slow variable $t_2$ and the slow-varying envelopes $f_{m}^{(1)}$ ($m=1,2$) is for cancelling such singularity, which is reflected by the
second term on the RHS of the above equation. Then by a solvablilty condition [i.e. cancelling the secular term in Eq.~(\ref{rho1 3rd Eqs})] yields closed equations for $f_{1}^{(1)}$ and $f_{2}^{(1)}$, which can be solved analytically~\cite{LiHJ}.

The general expression of $a_{21}^{(3)}$ are given by $a_{21}^{(3)}\equiv a_{21}^{(3),\rm LA}+N_{a}\int d^3{\bf r}^{\prime}V({\bf r}^{\prime}-{\bf r})a_{21}^{(3),\rm RR}$, contributed by two parts:
\begin{subequations}
\begin{align}
a_{21}^{(3),\rm LA}&=\sum_{m=1}^{2}v_{1m}g_{m}^{(3),\rm LA}e^{i\lambda_{m}t_{0}}+\sum_{l}w_{1l}^{\rm LA}e^{i\mu_{l}t_{0}}+a_{21}^{(3),\rm LA}(\infty),\\
a_{21}^{(3),\rm RR}&=\sum_{m=1}^{2}v_{1m}g_{m}^{(3),\rm RR}e^{i\lambda_{m}t_{0}}+\sum_{l}w_{1l}^{\rm RR}e^{i\mu_{l}t_{0}}+a_{21}^{(3),\rm RR}(\infty).
\end{align}
\label{rho1 3rd Sols}
\end{subequations}
Here $a_{21}^{(3),\rm LA}(\infty)$ and $a_{21}^{(3),\rm RR}(\infty)$ are corresponding steady-state solutions, given in Ref.~\cite{BaiOE}; $w_{1l}^{\rm LA(RR)}$ and $\mu_{l}$ are coefficients obtained when calculating the particular solutions stemming from the inhomogeneous terms; $g_{m}^{(3),\rm LA(RR)}$ are determined from the initial condition, given as
$g_{1}^{(3),\rm LA}=\{v_{22}[a_{21}^{(3),\rm LA(RR)}(0)-a_{21}^{(3),\rm LA(RR)}(\infty)-\sum_{l}w_{1l}^{\rm LA(RR)}]-v_{12}[a_{31}^{(3),\rm LA(RR)}(0)-a_{31}^{(3),\rm LA(RR)}(\infty)-\sum_{l}w_{1l}^{\rm LA(RR)}]\}/(v_{11}v_{22}-v_{21}v_{12})$,
$g_{2}^{(3),\rm LA}=\{v_{11}[a_{31}^{(3),\rm LA(RR)}(0)-a_{31}^{(3),\rm LA(RR)}(\infty)-\sum_{l}w_{1l}^{\rm LA(RR)}]-v_{21}[a_{21}^{(3),\rm LA(RR)}(0)-a_{21}^{(3),\rm LA(RR)}(\infty)-\sum_{l}w_{1l}^{\rm LA(RR)}]\}/(v_{11}v_{22}-v_{21}v_{12})$,
where $a_{21}^{(3),\rm LA}(0)=4\gamma_{21}/(\Gamma_{12}|d_{21}|^2)$, $a_{31}^{(3),\rm LA}(0)=a_{21}^{(3),\rm RR}(0)=a_{31}^{(3),\rm RR}(0)=0$.
%$a_{31}^{(3)}$ is not used and thus omitted here.
%

After returning to the original variables, we obtain
\begin{equation}
\rho_{21}(t)\approx a_{21}^{(1)}(t)\Omega_{p}+\left[a_{21}^{(3),\rm LA}(t)+N_{a}\int d^3{\bf r}'V({\bf r}'-{\bf r})a_{21}^{(3),\rm RR}({\bf r}'-{\bf r},t)\right]|\Omega_{p}|^2\Omega_{p},
\end{equation}
which is just Eq.~(\ref{rho21}) of the main context. Because the optical susceptibility of the system is given by $\chi_p(t)=[N_{a}|{\bf p}_{12}|^2/(\varepsilon_0 \hbar \Omega_p)] \rho_{21}(t)$, the transient optical response of the Rydberg-EIT can be described by the atomic coherence $\rho_{21}(t)$. The dispersion property of the system, described by ${\rm Re}(\rho_{21})$, is shown in Fig.~\ref{Fig6}(d) and Fig.~\ref{Fig6}(e), which is similar to the result solved with the two-atom model shown in Fig.~\ref{Fig6}(b) and Fig.~\ref{Fig6}(c).
%

%%%%%%%%%%%%%%%%%%%%%%%%%%%%%%%%%%%%%%%%%%%%%

\end{document}